\documentclass[prb,tighten,eqsecnum,showpacs]{revtex4}

\usepackage{amssymb,amsmath}

\usepackage{graphicx}
\newcommand{\dd}{\text{d}}

\begin{document}
\title{Mass transport in a strongly sheared binary mixture of Maxwell molecules}
\author{Vicente Garz\'o}
\email{vicenteg@unex.es} \homepage{http://www.unex.es/eweb/fisteor/vicente/} \affiliation{Departamento de
F\'{\i}sica, Universidad de Extremadura, E-06071 Badajoz, Spain}
\begin{abstract}
Transport coefficients associated with the mass flux of a binary mixture of Maxwell molecules under uniform
shear flow are exactly determined from the Boltzmann kinetic equation. A normal solution is obtained via a
Chapman--Enskog-like expansion around a local shear flow distribution that retains all the hydrodynamics orders
in the shear rate. In the first order of the expansion the mass flux is proportional to the gradients of mole
fraction, pressure, and temperature but, due to the anisotropy induced in the system by the shear flow, mutual
diffusion, pressure diffusion and thermal diffusion tensors are identified instead of the conventional scalar
coefficients. These tensors are obtained in terms of the shear rate and the parameters of the mixture (particle
masses, concentrations, and force constants). The description is made both in the absence and in the presence of
an external thermostat introduced in computer simulations to compensate for the viscous heating. As expected,
the analysis shows that there is not a simple relationship between the results with and without the thermostat.
The dependence of the three diffusion tensors on the shear rate is illustrated in the tracer limit case, the
results showing that the deviation of the generalized transport coefficients from their equilibrium forms is in
general quite important. Finally, the generalized transport coefficients associated with the momentum and heat
transport are evaluated from a model kinetic equation of the Boltzmann equation.
\end{abstract}

 \draft
\date{\today}
\maketitle

\section{Introduction}
\label{sec1}

The description of transport properties for states close to equilibrium in gaseous binary mixtures is well
established. In these situations, the Curie principle \cite{GM84} states that the presence of a velocity
gradient (second-rank tensorial quantity) cannot modify a vectorial quantity such as the mass flux ${\bf j}_1$,
which is generated by gradients of mole fraction $x_1$, pressure $p$, and temperature $T$. As a consequence, the
mutual diffusion coefficient $D$ (which couples the mass current with $\nabla x_1$), the pressure diffusion
coefficient $D_p$ (which couples the mass current with $\nabla p$) and the thermal diffusion coefficient $D_T$
(which couples the mass current with $\nabla T$) do not depend on the velocity gradient. However, when the shear
rate applied is large, non-Newtonian effects are important so that the Curie principle does not hold and the
coefficients associated with the mass transport are affected by the presence of shear flow. In particular, if
the spatial gradients $\nabla x_1$, $\nabla p$, and $\nabla T$ are weak, one expects that the flux ${\bf j}_1$
is still linear in these gradients but the standard scalar coefficients $\{D, D_{p},D_{T}\}$ must be replaced by
the shear-rate dependent second-rank tensors  $\{D_{ij}, D_{p,ij},D_{T,ij}\}$. The aim of this paper is to
determine the above tensors in the framework of the Boltzmann equation.

We are interested in a situation where {\em weak} spatial gradients of mole fraction, pressure, and temperature
coexist with a {\em strong} shear rate. Under these conditions, the application of the conventional
Chapman-Enskog expansion \cite{CC70} around the local equilibrium state to get higher order hydrodynamic effects
(Burnett, super-Burnett, $\ldots$) to the mass flux turns out to be extremely difficult. This gives rise to look
for alternative approaches. A possibility is to expand around a more relevant reference state than local
equilibrium. Since we want to compute the mass transport in a strongly sheared mixture, the so-called uniform
shear flow (USF) state can be chosen as the reference state. The USF state is characterized by constant mole
fractions, a uniform temperature, and a linear velocity profile $u_{x}=ay$, where $a$ is the constant shear
rate. Due to its simplicity, this state has been widely used in the past to shed light on the complexities
associated with the {\em nonlinear} response of the system to the action of strong shearing. In addition, the
USF state is one of the rare exceptions for which the hierarchy of moments of the Boltzmann equation admits an
exact solution for single \cite{TM80} and multicomponent gases \cite{MGS95} of Maxwell molecules (repulsive
potential of the form $r^{-4}$). In this case, explicit expressions of the pressure tensor (which is the
relevant irreversible flux of the problem) have been obtained for {\em arbitrary} values of the shear rate and
the parameters of the system (masses, concentrations and force constants).

As said before, here we want to compute the mass transport under USF for Maxwell molecules. Since the mixture is
slightly perturbed from the USF, the Boltzmann equation can be solved by an expansion in small gradients around
the (local) shear flow distribution instead of the (local) equilibrium. This is the main feature of the
expansion since the reference state is not restricted to small values of the shear rate. In the first order of
the expansion, the set of generalized transport coefficients $\{D_{ij}, D_{p,ij},D_{T,ij}\}$ are identified from
the mass flux ${\bf j}_1$ as nonlinear functions of the shear rate and the parameters of the mixture. This
Chapman-Enskog-like expansion has been used to analyze transport properties in spatially inhomogeneous states
near USF in the case of ordinary gases \cite{LD97,GS03} and more recently in the context of granular gases.
\cite{L06,G06,G07}

Some previous attempts have been carried out earlier by the author and coworkers \cite{GLH91, GLH92,
GLH95,MG98,MGLH00} in the case of the diffusion tensor $D_{ij}$. However, all these studies have been restricted
to perturbed {\em steady} states with the constraints $p=\text{const}$ and $T=\text{const}$. Although steady
states are in general desirable for practical purposes, especially in computer simulations, \cite{EM90} here we
extend the above studies to a general time and space dependence of the hydrodynamic fields. This allows us to
evaluate new contributions to the mass flux (those proportional to $\nabla p$ and $\nabla T$), which where not
taken into account in the previous studies. \cite{GS03}

The plan of the paper is as follows. First, a brief summary of the results obtained from the Boltzmann equation
for a binary mixture of Maxwell molecules under USF is presented in Sec.\ \ref{sec2}. Section \ref{sec3} deals
with the perturbation scheme used to solve the Boltzmann equation for the mixture to first order in the
deviations of the hydrodynamic field gradients from their values in the reference shear flow state. The
generalized transport coefficients characterizing the mass transport around USF are also defined in Sec.\
\ref{sec3}. These coefficients are explicitly obtained in Sec.\ \ref{sec4} with and without the presence of an
external thermostat introduced usually in computer simulations to compensate for the viscous heating. The
dependence of some of these coefficients on the shear rate is illustrated with detail in the tracer limit case,
showing that the influence of shear flow on mass transport is quite significant. The paper is closed by a brief
discussion of the results in Sec.\ \ref{sec5}, the generalized transport coefficients associated with the
momentum and heat transport are evaluated in Appendix \ref{appD} from a simple model kinetic equation of the
Boltzmann equation.

\section{A binary mixture under uniform shear flow}
\label{sec2}

We consider a {\em dilute} binary mixture where $f_s({\bf r},{\bf v};t)$ is the one-particle velocity
distribution function of species $s$ ($s=1,2$). The time evolution of the distributions $f_s$ is given by the
set of two coupled nonlinear Boltzmann equations:
\begin{equation}
\left( \partial _{t}+{\bf v}\cdot\nabla+ \frac{\partial}{\partial {\bf v}}\cdot \frac{{\bf F}_s}{m_s} \right)
f_{s}({\bf r},{\bf v},t)=\sum_{r=1}^2 J_{sr}\left[ {\bf v}|f_{s}(t),f_{r}(t)\right] \;, \label{2.1}
\end{equation}
where $m_s$ is the mass of a particle of species $s$, ${\bf F}_s$ is a possible external force acting on
particles of species $s$, and $J_{sr}\left[{\bf v}|f_{s},f_{r}\right]$ is the Boltzmann collision operator,
which in standard notation reads \cite{CC70}
\begin{equation}
\label{2.2} J_{sr}[f_s,f_r]=\int d{\bf v}_1 \int d\Omega |{\bf v}-{\bf v}_1|\sigma_{sr}({\bf v}-{\bf
v}_1,\theta) [f_s({\bf v}') f_r({\bf v}'_1)-f_s({\bf v}) f_r({\bf v}_1)] \;.
\end{equation}
The basic moments of $f_s$ are the species number densities
\begin{equation}
\label{2.3} n_s=\int \dd{\bf v} f_s,
\end{equation}
and the mean velocity of species $s$
\begin{equation}
\label{2.4} {\bf u}_s=\frac{1}{n_s}\int \dd{\bf v} {\bf v} f_s.
\end{equation}
These quantities define the total number density $n=\sum_s n_s$ and the flow velocity ${\bf
u}=\sum_s\;\rho_s{\bf u}_s/\rho$, where $\rho_s=m_sn_s$ is the mass density of species $s$ and $\rho=\sum_s\;
\rho_s$ is the total mass density. The temperature $T$ is defined as
\begin{equation}
\label{2.5} nk_BT=\sum_s\,n_sk_BT_s=\sum_s \frac{m_s}{3}\int \dd{\bf v} {\bf V}^2 f_s,
\end{equation}
where $k_B$ is the Boltzmann constant and ${\bf V}={\bf v}-{\bf u}$ is the peculiar velocity. The second
identity in (\ref{2.5}) defines the partial kinetic temperatures $T_s$ of species $s$. They measure the mean
kinetic energy of particles of species $s$. Moreover, in a dilute gas the hydrostatic pressure $p$ is given by
$p=nk_BT$. The quantities $n_s$, ${\bf u}$, and $T$ are associated with the densities of conserved quantities
(mass of each species, total momentum, and total energy). The corresponding balance equations define the
dissipative fluxes of mass
\begin{equation}
\label{2.6} {\bf j}_s=m_s\int \dd{\bf v}\; {\bf V} f_s,
\end{equation}
momentum (pressure tensor),
\begin{equation}
\label{2.7} {\sf P}=\sum_s\; {\sf P}_s=\sum_s\; m_s\int \dd{\bf v}\; {\bf V} {\bf V} f_s,
\end{equation}
and energy (heat flux)
\begin{equation}
\label{2.8} {\bf q}=\sum_s\;{\bf q}_s=\sum_s\; \frac{m_s}{2}\int \dd{\bf v}\; {\bf V}^2 {\bf V} f_s.
\end{equation}
The second equalities in Eqs.\ (\ref{2.7}) and (\ref{2.8}) define the partial contributions ${\sf P}_s$ and
${\bf q}_s$ to the pressure tensor and heat flux, respectively. The fact that the mass flux ${\bf j}_s$ is
defined with respect to the local center-of-mass velocity ${\bf u}$ implies that
\begin{equation}
\label{2.8.1} \sum_s\; {\bf j}_s={\bf 0}.
\end{equation}

The USF state is macroscopically defined by constant densities $n_s$, a spatially uniform temperature $T(t)$ and
a linear velocity profile ${\bf u}(y)={\bf u}_1(y)={\bf u}_2(y)=ay \widehat{{\bold x}}$, where $a$ is the {\em
constant} shear rate. Since $n_s$ and $T$ are uniform, then ${\bf j}_s={\bf q}={\bf 0}$, and the transport of
momentum (measured by the pressure tensor) is the relevant phenomenon. In the USF problem, the temperature tends
to increase in time due to viscous heating. Usually, an external force (thermostat) is introduced in computer
simulations to remove this heating effect and keep the temperature constant. \cite{EM90} The simplest choice is
a Gaussian isokinetic thermostat given by
\begin{equation}
\label{2.9} {\bf F}_s=-m_s \alpha {\bf V},
\end{equation}
where the thermostat parameter $\alpha$ is a function of the shear rate adjusted as to keep the temperature
constant. The implicit assumption behind the introduction of these forces is that they play a neutral role in
the transport properties, so that the latter are the same with and without a thermostat, when conveniently
scaled with the thermal speed. Nevertheless, this expectation is not in general true, except for some specific
situations and/or interaction potentials. \cite{DSBR86,GS03}

At a microscopic level, the USF is characterized by a velocity distribution function that becomes {\em uniform}
in the local Lagrangian frame, i.e., $f_s({\bf r},{\bf v};t)=f_s({\bf V},t)$. In that case, Eq.\ (\ref{2.1})
with the choice (\ref{2.9}) reduces to
\begin{equation}
\label{2.10} \frac{\partial}{\partial t}f_1-\frac{\partial}{\partial V_i}\left(a_{ij}V_j+\alpha
V_i\right)f_1=J_{11}[f_1,f_1]+J_{12}[f_1,f_2]
\end{equation}
and a similar equation for $f_2$. Here, $a_{ij}=a \delta_{ix}\delta_{jy}$. The hierarchy of velocity moments
associated with the Boltzmann equation (\ref{2.10}) can be recursively solved in the particular case of Maxwell
molecules, i.e., when particles of species $r$ and $s$ interact through a potential of the form
$V_{rs}(r)=\kappa_{rs}r^{-4}$. The key point is that for this interaction the collision rate $g \sigma_{rs}
(g,\theta)$ is independent of the relative velocity $g$ and so the collisional moments of order $k$ only involve
moments of degree smaller than or equal to $k$. In particular, the first- and second-degree collisional moments
are given by \cite{H66,GS67}
\begin{equation}
\label{2.10.1} m_s\int \; \dd{\bf v} {\bf V}J_{sr}[f_s,f_r]=-\frac{\lambda_{sr}}{m_sm_r}\left(\rho_s{\bf
j}_r-\rho_r{\bf j}_s\right),
\end{equation}
\begin{eqnarray}
\label{2.10.2} m_s\int \; \dd{\bf v} {\bf V}{\bf
V}J_{sr}[f_s,f_r]&=&\frac{\lambda_{sr}'}{(m_s+m_r)m_s}\left[\left(\rho_sp_r+\rho_rp_s-\frac{2}{3}{\bf j}_s\cdot
{\bf j}_r\right)\openone \right.\nonumber\\
& & \left.-\rho_s{\sf P}_r-\rho_r{\sf P}_s+{\bf j}_s{\bf j}_r+{\bf j}_r{\bf j}_s\right]\nonumber\\
& & -\frac{\lambda_{sr}}{(m_s+m_r)m_s}\left[2\left(\frac{m_s}{m_r}\rho_r{\sf P}_s-\rho_s{\sf P}_r\right)\right.
\nonumber\\
& & \left.+\left(1-\frac{m_s}{m_r}\right)({\bf j}_s{\bf j}_r+{\bf j}_r{\bf j}_s)\right],
\end{eqnarray}
where $p_s=\frac{1}{3}\text{tr}{\sf P}_s=n_sk_BT_s$ is the partial hydrostatic pressure and
\begin{equation}
\label{4.11} \lambda_{sr}=1.69\pi \left(\kappa_{sr}\frac{m_sm_r}{m_s+m_r}\right)^{1/2},\quad
\lambda_{sr}'=2.61\pi \left(\kappa_{sr}\frac{m_sm_r}{m_s+m_r}\right)^{1/2}.
\end{equation}
Thanks to the above property, exact expressions of the pressure tensor ${\sf P}$ for a binary mixture of Maxwell
molecules under USF were obtained some time ago. \cite{MGS95} The nonzero elements of ${\sf P}$ are related to
the rheological properties of the mixture, namely, the nonlinear shear viscosity and the viscometric functions.
In reduced units, they turn out to be nonlinear functions of the (reduced) shear rate $a^*=a/\zeta$ (where
$\zeta$ is a convenient time unit defined below) and the parameters of the mixture: the mass ratio
$\mu=m_1/m_2$, the mole fraction $x_1=n_1/n$ and the force constant ratios $\kappa_{11}/\kappa_{12}$ and
$\kappa_{22}/\kappa_{12}$. It must be noted that in the particular case of Maxwell molecules there is an exact
equivalence between the USF results with and without the external forces (\ref{2.9}). As will be shown below,
beyond the USF problem, the presence of the thermostat does not play a neutral role in the results and a certain
influence may exist.

\section{Chapman--Enskog-like expansion around USF}
\label{sec3}

As said in the Introduction, the main aim of this work is to analyze mass transport of a dilute binary mixture
subjected to USF. In that case, let us assume that the USF state is disturbed by small spatial perturbations.
The response of the system to those perturbations gives rise to contributions to the mass flux that can be
characterized by generalized transport coefficients. This Section is devoted to the evaluation of those
coefficients.

In order to analyze this problem we have to start from the set of Boltzmann equations (\ref{2.1}) with a general
time and space dependence. Let ${\bf u}_{0}={\sf a}\cdot {\bf r}$ be the flow velocity of the {\em undisturbed}
USF state, where the elements of the tensor ${\sf a}$ are $a_{ij}=a\delta_{ix}\delta_{jy}$. In the {\em
disturbed} state, however the true velocity ${\bf u}$ is in general different from ${\bf u}_0$, i.e., ${\bf
u}={\bf u}_0+\delta {\bf u}$, $\delta {\bf u}$ being a small perturbation to ${\bf u}_0$. As a consequence, the
true peculiar velocity is now ${\bf c}\equiv {\bf v}-{\bf u}={\bf V}-\delta{\bf u}$, where ${\bf V}={\bf v}-{\bf
u}_0$. In the Lagrangian frame moving with ${\bf u}_0$, the Boltzmann equations (\ref{2.1}) can be written as
\begin{subequations}
\begin{equation}
\label{3.1} \frac{\partial}{\partial t}f_1-\frac{\partial}{\partial V_i}\left(a_{ij}V_j+\alpha
V_i\right)f_1+\left({\bf V}+{\bf u}_0\right)\cdot \nabla f_1+\alpha \delta{\bf u}\cdot \frac{\partial
f_1}{\partial {\bf V}} =J_{11}[f_1,f_1]+J_{12}[f_1,f_2],
\end{equation}
\begin{equation}
\label{3.2} \frac{\partial}{\partial t}f_2-\frac{\partial}{\partial V_i}\left(a_{ij}V_j+\alpha
V_i\right)f_2+\left({\bf V}+{\bf u}_0\right)\cdot \nabla f_2+\alpha \delta{\bf u}\cdot \frac{\partial
f_2}{\partial {\bf V}} =J_{22}[f_2,f_2]+J_{21}[f_2,f_1],
\end{equation}
\end{subequations}
where here the derivative $\nabla f_s$ is taken at constant ${\bf V}$. In addition, in Eqs.\ (\ref{3.1}) and
(\ref{3.2}) the thermostat force has been assumed to be proportional to the actual peculiar velocity, ${\bf
F}_s=-m_s\alpha ({\bf V}-\delta {\bf u})$ where now the parameter $\alpha$ is in general a function of ${\bf r}$
and $t$ through their functional dependence on the hydrodynamic fields $n_s$ and $T$. The generalization of
$\alpha$ to the inhomogeneous case is essentially a matter of choice. Here, for the sake of simplicity, we will
take two different choices for $\alpha$: (i) $\alpha=0$, so that the temperature grows in time, and (ii) the
same expression obtained in the (pure) USF problem, except that the densities and temperature are replaced by
those of the general inhomogeneous state.

The macroscopic balance equations associated with this disturbed USF state are obtained by taking moments in
Eqs.\ (\ref{3.1}) and (\ref{3.2}) with the result
\begin{equation}
\label{3.3}
\partial_tn_s+{\bf u}_0\cdot \nabla n_s+\nabla \cdot (n_s\delta {\bf u})=-
\frac{\nabla \cdot {\bf j}_s}{m_s},
\end{equation}
\begin{equation}
\label{3.4}
\partial_t\delta u_i+a_{ij}\delta u_j+({\bf u}_0+\delta {\bf u})\cdot \nabla \delta u_i=-
\rho^{-1}\nabla_j P_{ij},
\end{equation}
\begin{equation}
\label{3.5} \frac{3}{2}n\partial_tT+\frac{3}{2}n({\bf u}_0+\delta {\bf u})\cdot \nabla
T=-aP_{xy}+\frac{3}{2}T\sum_{s=1}^2\frac{\nabla \cdot {\bf j}_s}{m_s}-\left(\nabla \cdot {\bf q}+{\sf P}:\nabla
\delta {\bf u}+3p\alpha\right),
\end{equation}
where the mass flux ${\bf j}_s$, the pressure tensor ${\sf P}$, and the heat flux ${\bf q}$ are defined by Eqs.\
(\ref{2.6}), (\ref{2.7}), and (\ref{2.8}), respectively, with the replacement ${\bf V}\rightarrow {\bf c}$. The
corresponding balance equations for the mole fraction $x_1=n_1/n$ and the pressure $p=nk_BT$ can be obtained
from Eqs.\ (\ref{3.3}) and (\ref{3.5}). They are given by
\begin{equation}
\label{3.5.1}
\partial_t x_1+({\bf u}_0+\delta {\bf u})\cdot \nabla x_1=-\frac{\rho}{n^2m_1m_2}\nabla \cdot {\bf j}_1,
\end{equation}
\begin{equation}
\label{3.5.2} \partial_tp+({\bf u}_0+\delta {\bf u})\cdot \nabla p+p\nabla \cdot \delta {\bf
u}=-\frac{2}{3}\left(aP_{xy}+\nabla \cdot {\bf q}+{\sf P}:\nabla \delta {\bf u}+3p\alpha\right).
\end{equation}

We assume that the deviations from the USF state are small. This means that the spatial gradients of the
hydrodynamic fields are small. For systems near equilibrium, the specific set of gradients contributing to each
flux is restricted by fluid symmetry, Onsager relations, and the form of entropy production. \cite{GM84}
However, in far from equilibrium situations (such as the one considered in this paper), only fluid symmetry
applies and so there is more flexibility in the representation of the heat and mass fluxes since they can be
defined in a variety of equivalent ways depending on the choice of hydrodynamic gradients used. In fact, some
care is required in comparing transport coefficients in different representations using different independent
gradients for the driving forces. Here, as in previous works, \cite{G07,GD02,GMD06} the mole fraction $x_1$, the
pressure $p$, the temperature $T$, and the local flow velocity $\delta {\bf u}$ are chosen as hydrodynamic
fields.

Since the system is strongly sheared, a solution to the set of Boltzmann equations (\ref{3.1}) and (\ref{3.2})
can be obtained by means of a generalization of the conventional Chapman-Enskog method \cite{CC70} in which the
velocity distribution function is expanded around a {\em local} shear flow reference state in terms of the small
spatial gradients of the hydrodynamic fields relative to those of USF. This is the main new ingredient of the
expansion. This type of Chapman-Enskog-like expansion has been already considered to get the set of shear-rate
dependent transport coefficients \cite{LD97,MG98,MGLH00} in thermostatted shear flow problems and it has also
been recently used  for inelastic gases. \cite{L06,G06,G07}

In the context of the Chapman--Enskog method, \cite{CC70} we look for a {\em normal} solution of the form
\begin{equation}
\label{3.7} f_s({\bf r}, {\bf V},t)\equiv f_s[A({\bf r}, t), {\bf V}],
\end{equation}
where
\begin{equation}
\label{3.6} A({\bf r},t)\equiv \{x_1({\bf r},t), p({\bf r}, t), T({\bf r}, t), \delta {\bf u}({\bf r},t)\}.
\end{equation}
This special solution expresses the fact that the space dependence of the reference shear flow is completely
absorbed in the relative velocity ${\bf V}$ and all other space and time dependence occurs entirely through a
{\em functional} dependence on the fields $A({\bf r}, t)$. The functional dependence (\ref{3.5}) can be made
local by an expansion of the distribution function in powers of the hydrodynamic gradients:
\begin{equation}
\label{3.8} f_s[A({\bf r}, t, {\bf V}] =f_s^{(0)}({\bf V})+ f_s^{(1)}({\bf V})+\cdots,
\end{equation}
where the reference zeroth-order distribution function corresponds to the USF distribution function but taking
into account the local dependence of the concentration, pressure and temperature and the change ${\bf
V}\rightarrow {\bf V}-\delta{\bf u}({\bf r}, t)={\bf c}$. The successive approximations $f_s^{(k)}$ are of order
$k$ in the gradients of $x_1$, $p$, $T$, and $\delta {\bf u}$ but retain all the orders in the shear rate $a$.
Here, only the first-order approximation will be analyzed.

When the expansion (\ref{3.8}) is substituted into the definitions (\ref{2.6}), (\ref{2.7}), and (\ref{2.8}),
one gets the corresponding expansions for the fluxes:
\begin{subequations}
\begin{equation}
\label{3.9} {\bf j}_s={\bf j}_s^{(0)}+{\bf j}_s^{(1)}+\cdots,
\end{equation}
\begin{equation}
\label{3.10} {\sf P}={\sf P}^{(0)}+{\sf P}^{(1)}+\cdots, \quad {\bf q}={\bf q}^{(0)}+{\bf q}^{(1)}+\cdots.
\end{equation}
\end{subequations}
Finally, as in the usual Chapman-Enskog method, the time derivative is also expanded as
\begin{equation}
\label{3.11}
\partial_t=\partial_t^{(0)}+\partial_t^{(1)}+\partial_t^{(2)}+\cdots,
\end{equation}
where the action of each operator $\partial_t^{(k)}$ is obtained from the hydrodynamic equations
(\ref{3.3})--(\ref{3.5}). These results provide the basis for generating the Chapman-Enskog solution to the
Boltzmann equations (\ref{3.1}) and (\ref{3.2}).

\subsection{Zeroth-order approximation}

Substituting the expansions (\ref{3.9})--(\ref{3.11}) into Eq.\ (\ref{3.1}), the kinetic equation for
$f_1^{(0)}$ is given by
\begin{equation}
\label{3.12} \frac{\partial}{\partial t}f_1^{(0)}-\frac{\partial}{\partial V_i}\left(a_{ij}V_j+\alpha
V_i\right)f_1^{(0)}+\left({\bf V}+{\bf u}_0\right)\cdot \nabla f_1^{(0)}+\alpha \delta{\bf u}\cdot
\frac{\partial f_1^{(0)}}{\partial {\bf V}} =J_{11}[f_1^{(0)},f_1^{(0)}]+J_{12}[f_1^{(0)},f_2^{(0)}].
\end{equation}
To lowest order in the expansion the conservation laws give
\begin{equation}
\label{3.13}
\partial_t^{(0)}x_1=0,\quad
T^{-1}\partial_t^{(0)}T=p^{-1}\partial_t^{(0)}p= -\frac{2}{3p}a P_{xy}^{(0)}-2\alpha,
\end{equation}
\begin{equation}
\label{3.14}
\partial_t^{(0)}\delta u_i+a_{ij} \delta u_j=0.
\end{equation}
If $\alpha=0$, then $T^{-1}\partial_t^{(0)}T=p^{-1}\partial_t^{(0)}p=-2aP_{xy}^{(0)}/3p$ while if
$\alpha=-aP_{xy}^{(0)}/3p$ then $\partial_t^{(0)}T=\partial_t^{(0)}p=0$.

Since $f_1^{(0)}$ is a normal solution, the time derivative in Eq.\ (\ref{3.12}) can be represented more
usefully as
\begin{eqnarray}
\label{3.15}
\partial_t^{(0)}f_1^{(0)}&=&\frac{\partial f_1^{(0)}}{\partial
x_1}\partial_t^{(0)} x_1+\frac{\partial f_1^{(0)}}{\partial p}\partial_t^{(0)} p+\frac{\partial
f_1^{(0)}}{\partial T}\partial_t^{(0)} T+\frac{\partial f_1^{(0)}}{\partial \delta
u_i}\partial_t^{(0)} \delta u_i\nonumber\\
&=&-\left(\frac{2}{3 p}a P_{xy}^{(0)}+2\alpha\right)\left(p\frac{\partial}{\partial p}+T\frac{\partial}{\partial
T}\right)f_1^{(0)}-a_{ij}\delta u_j
\frac{\partial}{\partial \delta u_i}f_1^{(0)}\nonumber\\
&=&-\left(\frac{2}{3 p}a P_{xy}^{(0)}+2\alpha\right)\left(p\frac{\partial}{\partial p}+T\frac{\partial}{\partial
T}\right)f_1^{(0)}+a_{ij}\delta u_j \frac{\partial}{\partial c_i}f_1^{(0)},
\end{eqnarray}
where in the last step we have taken into account that $f_1^{(0)}$ depends on $\delta {\bf u}$ only through the
peculiar velocity ${\bf c}$. Substituting Eq.\ (\ref{3.15}) into Eq.\ (\ref{3.12}) yields the following kinetic
equation for $f_1^{(0)}$:
\begin{eqnarray}
\label{3.16} -\left(\frac{2}{3p}a P_{xy}^{(0)}+2\alpha\right)& & \left(p\frac{\partial}{\partial
p}+T\frac{\partial}{\partial T}\right)f_1^{(0)} -ac_y\frac{\partial}{\partial c_x}f_1^{(0)} -\alpha
\frac{\partial }{\partial {\bf c}}\cdot \left({\bf
c}f_1^{(0)}\right)\nonumber\\
& &=J_{11}[f_1^{(0)},f_1^{(0)}]+J_{12}[f_1^{(0)},f_2^{(0)}].
\end{eqnarray}
A similar equation holds for $f_2^{(0)}$. The partial pressure tensors ${\sf P}_1^{(0)}$ and ${\sf P}_2^{(0)}$
can be obtained from Eq.\ (\ref{3.16}) and its counterpart for $f_2^{(0)}$ when one multiplies both equations by
$m_s {\bf c}{\bf c}$ and integrate over ${\bf c}$. Their explicit forms can be found in the Appendix of Ref.\
\onlinecite{MGS95}.

\subsection{First-order approximation}

The analysis to first order in the gradients is worked out in Appendix \ref{appA}. The distribution function
$f_1^{(1)}$ is of the form
\begin{equation}
\label{3.17} f_1^{(1)}={\boldsymbol {\cal A}}_{1}\cdot \nabla x_1+ {\boldsymbol {\cal B}}_{1}\cdot \nabla
p+{\boldsymbol {\cal C}}_{1}\cdot \nabla T+{\sf {\cal D}}_{1}:\nabla \delta {\bf u},
\end{equation}
where the vectors $\{{\boldsymbol {\cal A}}_{1}, {\boldsymbol {\cal B}}_{1}, {\boldsymbol {\cal C}}_{1}\}$, and
the tensor ${\sf {\cal D}}_{1}$ are functions of the true peculiar velocity ${\bf c}$. They are the solutions of
the following set of linear integral equations:
\begin{eqnarray}
\label{3.18}  & &-\left(\frac{2}{3 p}a
P_{xy}^{(0)}+2\alpha\right)\left(p\partial_p+T\partial_T\right){\boldsymbol {\cal A}}_{1}- \left(a
c_y\frac{\partial}{\partial c_x}+\alpha \frac{\partial }{\partial {\bf c}}\cdot {\bf c}\right){\boldsymbol
{\cal A}}_{1}+{\cal L}_1{\boldsymbol {\cal A}}_{1}+{\cal M}_1{\boldsymbol {\cal A}}_{2}\nonumber\\
& & ={\bf A}_{1}+\left(\frac{2a}{3p}\partial_{x_1}P_{xy}^{(0)}+2\partial_{x_1}\alpha \right)\left(p{\boldsymbol
{\cal B}}_{1}+T{\boldsymbol {\cal C}}_1\right),
\end{eqnarray}
\begin{eqnarray}
\label{3.19}  & &-\left(\frac{2}{3 p}a
P_{xy}^{(0)}+2\alpha\right)\left(p\partial_p+T\partial_T\right){\boldsymbol {\cal B}}_{1}- \left(a
c_y\frac{\partial}{\partial c_x}+\alpha \frac{\partial }{\partial {\bf c}}\cdot {\bf c}\right){\boldsymbol
{\cal B}}_{1}+{\cal L}_1{\boldsymbol {\cal B}}_{1}+{\cal M}_1{\boldsymbol {\cal B}}_{2}\nonumber\\
& & -\left[\frac{2a}{3}\partial_pP_{xy}^{(0)}+2(1+p\partial_p)\alpha\right]{\boldsymbol {\cal B}}_{1}={\bf
B}_{1}-\left[\frac{2aT}{3p^2}(1-p\partial_{p})P_{xy}^{(0)}-2T\partial_{p}\alpha \right]{\boldsymbol {\cal C}}_1,
\end{eqnarray}
\begin{eqnarray}
\label{3.20}  & &-\left(\frac{2}{3 p}a
P_{xy}^{(0)}+2\alpha\right)\left(p\partial_p+T\partial_T\right){\boldsymbol {\cal C}}_{1}- \left(a
c_y\frac{\partial}{\partial c_x}+\alpha \frac{\partial }{\partial {\bf c}}\cdot {\bf c}\right){\boldsymbol
{\cal C}}_{1}+{\cal L}_1{\boldsymbol {\cal C}}_{1}+{\cal M}_1{\boldsymbol {\cal C}}_{2}\nonumber\\
& & -\left[\frac{2a}{3p}(1+T\partial_T)P_{xy}^{(0)}+2(1+T\partial_T)\alpha\right]{\boldsymbol {\cal C}}_{1}={\bf
C}_{1}+\left(\frac{2a}{3}\partial_{T}P_{xy}^{(0)}+2p\partial_{T}\alpha \right){\boldsymbol {\cal B}}_1,
\end{eqnarray}
\begin{eqnarray}
\label{3.21}& & -\left(\frac{2}{3 p}a P_{xy}^{(0)}+2\alpha\right)\left(p\partial_p+T\partial_T\right){\cal
D}_{1,\ell j} - \left(a c_y\frac{\partial}{\partial c_x}+\alpha \frac{\partial }{\partial {\bf c}}\cdot {\bf
c}\right){\cal D}_{1,\ell j}-a\delta_{\ell y}{\cal D}_{1,xj}\nonumber\\
& & +{\cal L}_1{\cal D}_{1,\ell j}+{\cal M}_1{\cal D}_{2,\ell j}=D_{1,\ell j},
\end{eqnarray}
where ${\bf A}_{1}({\bf c})$, ${\bf B}_{1}({\bf c})$, ${\bf C}_{1}({\bf c})$, and ${\sf D}_1({\bf c})$ are
defined by Eqs.\ (\ref{a8})--(\ref{a11}), respectively. In addition, ${\cal L}_1$ and ${\cal M}_1$ are the
linearized Boltzmann collision operators around the reference USF state:
\begin{subequations}
\begin{equation}
\label{3.22} {\cal L}_1X=-\left(J_{11}[f_1^{(0)},X]+J_{11}[X,f_1^{(0)}]+J_{12}[X,f_2^{(0)}]\right),
\end{equation}
\begin{equation}
\label{3.23} {\cal M}_1X=-J_{12}[f_2^{(0)},X].
\end{equation}
\end{subequations}

In this paper we are mainly interested in evaluating the first-order contribution to the mass flux ${\bf
j}_1^{(1)}$. It is defined as
\begin{equation}
{\bf j}_{1}^{(1)}=m_{1}\int \dd{\bf c}\,{\bf c}\,f_1^{(1)}, \quad {\bf j}_{2}^{(1)}=-{\bf j}_{1}^{(1)}.
\label{3.24}
\end{equation}
Use of Eq.\ (\ref{3.17}) into Eq.\ (\ref{3.24}) gives the expression
\begin{equation}
\label{3.25} j_{1,i}^{(1)}=-\frac{m_1m_2n}{\rho}D_{ij}\frac{\partial x_1}{\partial
r_j}-\frac{\rho}{p}D_{p,ij}\frac{\partial p}{\partial r_j}-\frac{\rho}{T}D_{T,ij}\frac{\partial T}{\partial
r_j},
\end{equation}
where
\begin{equation}
\label{3.26} D_{ij}=-\frac{\rho}{m_2n}\int \dd{\bf c}\,c_i\;{\cal A}_{1,j}({\bf c}),
\end{equation}
\begin{equation}
\label{3.27} D_{p,ij}=-\frac{m_1p}{\rho}\int \dd{\bf c}\,c_i\;{\cal B}_{1,j}({\bf c}),
\end{equation}
\begin{equation}
\label{3.28} D_{T,ij}=-\frac{m_1T}{\rho}\int \dd{\bf c}\,c_i\;{\cal C}_{1,j}({\bf c}).
\end{equation}
Upon writing Eqs.\ (\ref{3.26})--(\ref{3.28}) use has been made of the symmetry properties of ${\boldsymbol
{\cal A}}_{1}$, ${\boldsymbol {\cal B}}_{1}$, and ${\boldsymbol {\cal C}}_{1}$. In general, the set of {\em
generalized} transport coefficients $D_{ij}$, $D_{p,ij}$, and $D_{T,ij}$ are nonlinear functions of the shear
rate and the parameters of the mixture. It is apparent that the anisotropy induced by the presence of shear flow
gives rise to new transport coefficients for the mass flux, reflecting broken symmetry. According to Eq.\
(\ref{3.25}), the mass flux is expressed in terms of a diffusion tensor $D_{ij}$, a pressure diffusion tensor
$D_{p,ij}$, and a thermal diffusion tensor $D_{T,ij}$.

To get the explicit dependence of the above transport coefficients on the parameter space of the problem, the
form of $\alpha$ must be chosen. As said before, two choices will be considered here: (i) the unthermostatted
case $\alpha=0$, and (ii) the thermostatted case $\alpha=-aP_{xy}^{(0)}/3p$. Both cases will be separately
studied in the next Section.

\section{Mass transport under shear flow}
\label{sec4}

This Section is devoted to the determination of the generalized transport coefficients $D_{ij}$, $D_{p,ij}$, and
$D_{T,ij}$ associated with the mass transport for the two choices of the thermostat parameter. These
coefficients are given in terms of the solutions to the integral equations (\ref{3.18})--(\ref{3.20}).

\subsection{Unthermostatted USF state}

In the absence of an external thermostat ($\alpha=0$), the integral equations (\ref{3.18})--(\ref{3.20}) become
\begin{eqnarray}
\label{4.1} & & -\frac{2}{3 p}a P_{xy}^{(0)}\left(p\partial_p+T\partial_T\right){\boldsymbol {\cal A}}_{1}- a
c_y\frac{\partial}{\partial c_x}{\boldsymbol {\cal A}}_{1}+{\cal L}_1{\boldsymbol {\cal A}}_{1}+{\cal
M}_1{\boldsymbol {\cal A}}_{2}\nonumber\\
& & ={\bf A}_{1}+\frac{2a}{3p}\left(p{\boldsymbol {\cal B}}_{1}+T{\boldsymbol {\cal
C}}_1\right)(\partial_{x_1}P_{xy}^{(0)}),
\end{eqnarray}
\begin{eqnarray}
\label{4.2}& & -\frac{2}{3 p}a P_{xy}^{(0)}\left(p\partial_p+T\partial_T\right){\boldsymbol {\cal B}}_{1}-
\left(\frac{2a}{3}\partial_pP_{xy}^{(0)}+a c_y\frac{\partial}{\partial c_x}\right){\boldsymbol {\cal
B}}_{1}+{\cal L}_1{\boldsymbol {\cal B}}_{1}+{\cal M}_1{\boldsymbol {\cal B}}_{2}\nonumber\\
& & ={\bf B}_{1}-\frac{2aT}{3p^2}{\boldsymbol {\cal C}}_1(1-p\partial_{p})P_{xy}^{(0)},
\end{eqnarray}
\begin{eqnarray}
\label{4.3} & &-\frac{2}{3 p}a P_{xy}^{(0)}\left(p\partial_p+T\partial_T\right){\boldsymbol {\cal C}}_{1}-
\left[\frac{2a}{3p}(1+T\partial_T)P_{xy}^{(0)}+a c_y\frac{\partial}{\partial c_x}\right]{\boldsymbol {\cal
C}}_{1}+{\cal L}_1{\boldsymbol {\cal C}}_{1}+{\cal M}_1{\boldsymbol {\cal C}}_{2}\nonumber\\
& & ={\bf C}_{1}+\frac{2a}{3}{\boldsymbol {\cal B}}_1(\partial_{T}P_{xy}^{(0)}).
\end{eqnarray}
The dependence of $P_{ij}^{(0)}$ on the pressure $p$ and temperature $T$ occurs explicitly and through its
dependence on the reduced shear rate $a^*=a/\zeta$. Here, the effective collision frequency $\zeta$ is given by
\cite{MGS95}
\begin{equation}
\label{4.4} \zeta=2n\frac{\lambda_{12}'}{m_1+m_2}=2\frac{p}{k_BT}\frac{\lambda_{12}'}{m_1+m_2},
\end{equation}
where $\lambda_{12}'$ is defined in Eq.\ (\ref{4.11}). Consequently,
\begin{equation}
\label{4.5} \partial_p P_{ij}^{(0)}=\partial_p p P_{ij}^*(a^*)=\left(1-a^*\frac{\partial} {\partial
a^*}\right)P_{ij}^*(a^*),
\end{equation}
\begin{equation}
\label{4.6} \partial_T P_{ij}^{(0)}=\partial_T p P_{ij}^*(a^*)=\frac{p}{T} a^*\frac{\partial} {\partial
a^*}P_{ij}^*(a^*),
\end{equation}
where $P_{ij}^*=P_{ij}^{(0)}/p$. In addition, the dependence of $P_{ij}^{(0)}$ on the mole fraction $x_1$ is
also rather intricate and so the derivatives with respect to $x_1$ must be carried out with care. \cite{MG98}
The generalized coefficients $D_{ij}$, $D_{p,ij}$, and $D_{T,ij}$ can be obtained from Eqs.\
(\ref{4.1})--(\ref{4.3}) when one multiplies those equations by $m_1 c_i$ and integrates over ${\bf c}$. After
some algebra (some technical details are provided in Appendix \ref{appC}), one arrives at the following set of
coupled algebraic equations:
\begin{eqnarray}
\label{4.23} \left[\left(\frac{\rho\lambda_{12}}{m_1m_2}- \frac{2}{3}a
P_{xy}^*\right)\delta_{ik}+a_{ik}\right]D_{kj}&=&\frac{\rho k_BT}{m_1m_2}\left(\partial_{x_1}P_{1,ij}^{*}-
\frac{\rho_1}{\rho}\partial_{x_1}P_{ij}^*\right)\nonumber\\
& &+\frac{2a\rho^2}{3m_1m_2n}(\partial_{x_1}P_{xy}^{*}) \left(D_{p,ij}+D_{T,ij}\right),
\end{eqnarray}
\begin{eqnarray}
\label{4.24} \left[\left(\frac{\rho\lambda_{12}}{m_1m_2}-\frac{2a}{3p}(1-a^*\partial_{a^*})P_{xy}^*\right)
\delta_{ik}+a_{ik}\right]D_{p,kj}&=& \frac{p}{\rho}(1-a^*\partial_{a^*})
\left(P_{1,ij}^*-\frac{\rho_1}{\rho}P_{ij}^*\right)\nonumber\\
& & -\frac{2a}{3}a^*D_{T,ij}(\partial_{a^*}P_{xy}^*),
\end{eqnarray}
\begin{eqnarray}
\label{4.25} \left[\left(\frac{\rho\lambda_{12}}{m_1m_2}-\frac{2a}{3}(1+a^*\partial_{a^*})P_{xy}^*\right)
\delta_{ik}+a_{ik}\right]D_{T,kj}&=& \frac{p}{\rho}a^*\partial_{a^*}
\left(P_{1,ij}^*-\frac{\rho_1}{\rho}P_{ij}^*\right)\nonumber\\
& & +\frac{2a}{3}a^*D_{p,ij}(\partial_{a^*}P_{xy}^*),
\end{eqnarray}
where ${\sf P}_s^*={\sf P}_s^{(0)}/p$ and use has been made of the relations (\ref{4.5}) and (\ref{4.6}).

In the absence of shear field ($a=0$), then $P_{s,ij}^*=x_s\delta_{ij}$, and $P_{ij}^*=\delta_{ij}$, so that
Eqs.\ (\ref{4.23})--(\ref{4.25}) have the solutions $D_{ij}=D_0\delta_{ij}$, $D_{p,ij}=D_{p,0}\delta_{ij}$, and
$D_{T,ij}=0$, where $D_0$ and $D_{p,0}$ are the conventional Navier-Stokes transport coefficients for Maxwell
molecules. \cite{CC70} Their expressions are
\begin{equation}
\label{4.26} D_0=\frac{k_BT}{\lambda_{12}},\quad D_{p,0}=\frac{\rho_1\rho_2}{\rho^3}(m_2-m_1)D_0.
\end{equation}
In this case the mass flux ${\bf j}^{(1)}$ can be written as \cite{GM84,CC70}
\begin{equation}
\label{4.26.1} {\bf j}_1^{(1)}=-\frac{m_1m_2\rho_1\rho_2}{k_B\rho^2}D_0\frac{(\nabla\phi_1)_T-
(\nabla\phi_2)_T}{T},
\end{equation}
where
\begin{equation}
\label{4.26.2} \left(\frac{\nabla \phi_s}{T}\right)_T=\frac{1}{m_s}\nabla \ln (x_sp),
\end{equation}
$\phi_s$ being the chemical potential per unit mass. The fact that the thermal diffusion coefficient vanishes
when $a^*=0$ is due to the interaction potential considered (Maxwell molecules) since this coefficient is
different from zero for more general interaction potentials. \cite{CC70} However, when the mixture is strongly
sheared, the Boltzmann equation leads to contributions to the mass flux proportional to the thermal gradient,
even for Maxwell molecules.

In the case of mechanically equivalent particles ($\mu=1$, $\kappa_{11}=\kappa_{22}=\kappa_{12}$),
$P_{1,ij}^*=x_1P_{ij}^*$, $\partial_{x_1}P_{1,ij}^{(0)}=P_{1,ij}^{(0)}/x_1=P_{xy}^{(0)}$, and so
$D_{p,ij}=D_{T,ij}=0$. Moreover, Eq.\ (\ref{4.23}) reduces to
\begin{equation}
\label{4.26.3} D_{ij}=\frac{m^{-1}}{n\lambda_{12}/m-\frac{2}{3}aP_{xy}^*}\left(\delta_{ik}-
\frac{a_{ik}}{n\lambda_{12}/m-\frac{2}{3}aP_{xy}^*}\right)P_{kj}^{(0)}.
\end{equation}
Equation (\ref{4.26.3}) is consistent with previous results derived for the self-diffusion tensor.
\cite{MD83,GSB90} Furthermore, known results for the diffusion tensor \cite{GLH91} are also recovered in the
tracer limit ($x_1\to 0$).

\begin{figure}
\includegraphics[width=0.6 \columnwidth,angle=0]{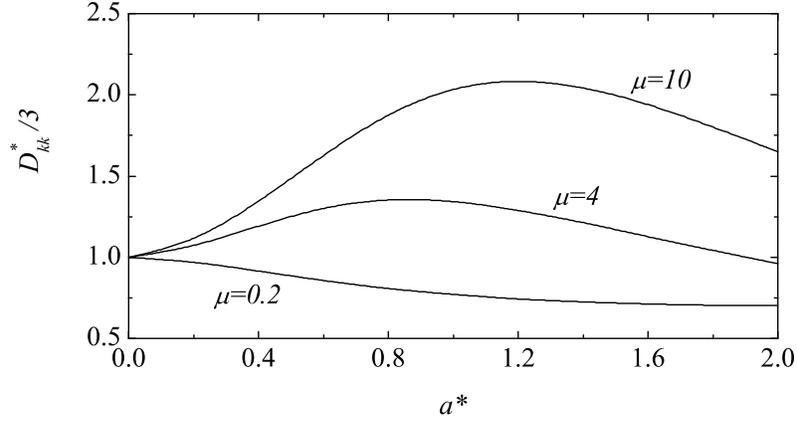}
\caption{Shear-rate dependence of the trace $\frac{1}{3}D_{kk}^*$ of the mutual diffusion tensor for $x_1=0$,
$\kappa_{22}=\kappa_{12}$ and several values of the mass ratio $\mu=m_1/m_2$. \label{fig1}}
\end{figure}
\begin{figure}
\includegraphics[width=0.6 \columnwidth,angle=0]{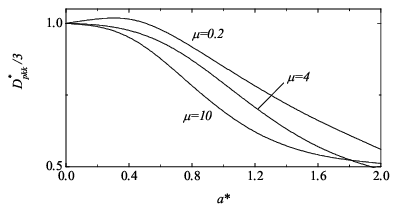}
\caption{Shear-rate dependence of the trace $\frac{1}{3}D_{p,kk}^*$ of the pressure diffusion tensor for
$x_1=0$, $\kappa_{22}=\kappa_{12}$ and several values of the mass ratio $\mu=m_1/m_2$. \label{fig2}}
\end{figure}
\begin{figure}
\includegraphics[width=0.6 \columnwidth,angle=0]{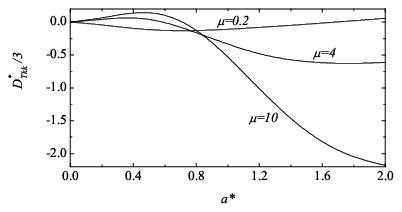}
\caption{Shear-rate dependence of the trace $\frac{1}{3}D_{T,kk}^*$ of the thermal diffusion tensor for $x_1=0$,
$\kappa_{22}=\kappa_{12}$ and several values of the mass ratio $\mu=m_1/m_2$. \label{fig3}}
\end{figure}

\subsection{Thermostatted USF state}

Let us assume now that an external thermostat is introduced to compensate for the viscous heating effect. In
this case, $\alpha=-aP_{xy}^{(0)}/3p$, and the integral equations (\ref{3.18})--(\ref{3.20}) become
\begin{equation}
\label{4.27} -\left(a c_y\frac{\partial}{\partial c_x}+\alpha \frac{\partial }{\partial {\bf c}}\cdot {\bf
c}\right){\boldsymbol {\cal A}}_{1}+{\cal L}_1{\boldsymbol {\cal A}}_{1}+{\cal M}_1{\boldsymbol {\cal
A}}_{2}={\bf A}_{1},
\end{equation}
\begin{equation}
\label{4.28} -\left(a c_y\frac{\partial}{\partial c_x}+\alpha \frac{\partial }{\partial {\bf c}}\cdot {\bf
c}\right){\boldsymbol {\cal B}}_{1}+{\cal L}_1{\boldsymbol {\cal B}}_{1}+{\cal M}_1{\boldsymbol {\cal
B}}_{2}={\bf B}_{1},
\end{equation}
\begin{equation}
\label{4.29} -\left(a c_y\frac{\partial}{\partial c_x}+\alpha \frac{\partial }{\partial {\bf c}}\cdot {\bf
c}\right){\boldsymbol {\cal C}}_{1}+{\cal L}_1{\boldsymbol {\cal C}}_{1}+{\cal M}_1{\boldsymbol {\cal
C}}_{2}={\bf C}_{1}.
\end{equation}
\begin{figure}
\includegraphics[width=0.6 \columnwidth,angle=0]{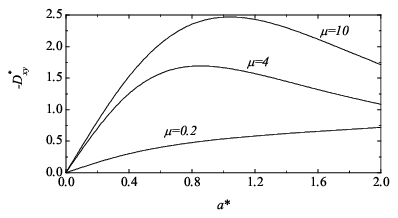}
\caption{Shear-rate dependence of the off-diagonal element $-D_{xy}^*$ of the mutual diffusion tensor for
$x_1=0$, $\kappa_{22}=\kappa_{12}$ and several values of the mass ratio $\mu=m_1/m_2$. \label{fig4}}
\end{figure}
\begin{figure}
\includegraphics[width=0.6 \columnwidth,angle=0]{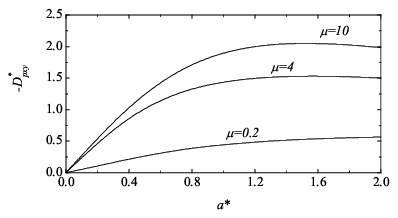}
\caption{Shear-rate dependence of the off-diagonal element $-D_{p,xy}^*$ of the pressure diffusion tensor for
$x_1=0$, $\kappa_{22}=\kappa_{12}$ and several values of the mass ratio $\mu=m_1/m_2$. \label{fig5}}
\end{figure}
\begin{figure}
\includegraphics[width=0.6 \columnwidth,angle=0]{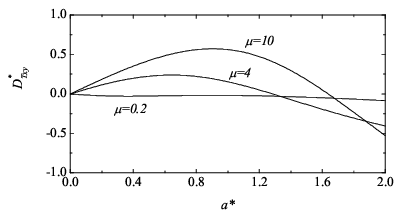}
\caption{Shear-rate dependence of the off-diagonal element $D_{T,xy}^*$ of the thermal diffusion tensor for
$x_1=0$, $\kappa_{22}=\kappa_{12}$ and several values of the mass ratio $\mu=m_1/m_2$. \label{fig6}}
\end{figure}
In contrast to what happens in the unthermostatted case, the different integral equations are now decoupled and
hence the generalized coefficients of the mass transport can be obtained more easily. The mathematical steps to
get them are similar to those made before when $\alpha=0$ and so only the final results are presented. The
explicit expressions for $D_{ij}$, $D_{p,ij}$, and $D_{T,ij}$ are given by
\begin{equation}
\label{4.30} D_{ij}=\frac{\rho k_BT}{m_1m_2}\frac{1}{\alpha+\frac{\rho\lambda_{12}}{m_1m_2}}\left(\delta_{ik}-
\frac{a_{ik}}{\alpha+\frac{\rho\lambda_{12}}{m_1m_2}}\right)\left(\partial_{x_1}P_{1,kj}^{*}-
\frac{\rho_1}{\rho}\partial_{x_1}P_{kj}^*\right),
\end{equation}
\begin{equation}
\label{4.31} D_{p,ij}=\frac{p}{\rho}\frac{1}{\alpha+\frac{\rho\lambda_{12}}{m_1m_2}}\left(\delta_{ik}-
\frac{a_{ik}}{\alpha+\frac{\rho\lambda_{12}}{m_1m_2}}\right)(1-a^*\partial_{a^*})
\left(P_{1,kj}^*-\frac{\rho_1}{\rho}P_{kj}^*\right),
\end{equation}
\begin{equation}
\label{4.32} D_{T,ij}=\frac{p}{\rho}\frac{a^*}{\alpha+\frac{\rho\lambda_{12}}{m_1m_2}}\left(\delta_{ik}-
\frac{a_{ik}}{\alpha+\frac{\rho\lambda_{12}}{m_1m_2}}\right)\partial_{a^*}
\left(P_{1,kj}^*-\frac{\rho_1}{\rho}P_{kj}^*\right).
\end{equation}
In order to get these expressions use has been made of the identity
\begin{equation}
\label{c1} \left(b\openone+{\sf a}\right)^{-1}=b^{-1}\openone-b^{-2}{\sf a},
\end{equation}
where $b$ is an arbitrary constant and ${\sf a}$ is the tensor with elements $a_{ij}=a\delta_{ix}\delta_{jy}$.

In the case of mechanically equivalent particles, $D_{p,ij}=D_{T,ij}=0$ and Eq.\ (\ref{4.30}) reduces to
\begin{equation}
\label{4.33} D_{ij}=\frac{m^{-1}}{\alpha+n\lambda_{12}/m}\left(\delta_{ik}-
\frac{a_{ik}}{\alpha+n\lambda_{12}/m}\right)P_{kj}^{(0)}.
\end{equation}
Equation (\ref{4.33}) gives the self-diffusion tensor of tagged particles under thermostatted USF. \cite{GSB90}
For a general binary mixture, the expression (\ref{4.30}) for the diffusion tensor $D_{ij}$ coincides with the
one derived before \cite{MG98} in a stationary state with the constraints $p=\text{const}$ and $T=\text{const}$.
Finally, it is also apparent that, except for vanishing shear rates, the expressions of the generalized
transport coefficients (\ref{4.30})--(\ref{4.32}) in the thermostatted state differ from the ones derived in the
absence of a thermostat, Eqs.\ (\ref{4.23})--(\ref{4.25}). This shows again that the presence of the thermostat
affects the transport properties of the system.

\section{Illustrative examples in the tracer limit}
\label{sec5}

The results obtained in the preceding Section give all the relevant information on the influence of shear flow
on the mass transport. In general, the elements $D_{ij}$, $D_{p,ij}$ and $D_{T,ij}$ present a complex dependence
on the shear rate and the parameters of the mixture without any restriction on their values. However, although
the solution to Eqs.\ (\ref{4.23})--(\ref{4.25}) (in the unthermostatted case) and Eqs.\
(\ref{4.30})--(\ref{4.32}) (in the thermostatted case) is simple, it involves quite a tedious algebra due to the
complex dependence of the partial pressure tensors ${\sf P}_{s,ij}^{(0)}$ and the thermostat parameter $\alpha$
on the mole fraction $x_1$ and the reduced shear rate $a^*$. To show the shear-rate dependence of the tensors
$T_{ij}\equiv \left\{D_{ij}, D_{p,ij}, D_{T,ij}\right\}$ in a clearer way, the tracer limit ($x_1\to 0$) will be
considered here in detail. In addition, to make some contact with computer simulation results, the thermostatted
case will be studied. In the tracer limit case, ${\sf P}\simeq {\sf P}_2$ and the partial pressure tensors ${\sf
P}_1$ and ${\sf P}_2$ have a more simplified forms. \cite{GLH91} In particular, $\partial_{x_1}P_{ij}^{(0)}=0$
and $\partial_{x_1}P_{1,ij}^{(0)}=P_{1,ij}^{(0)}/x_1$. The explicit expressions of the partial pressure tensors
in the tracer limit are provided in Appendix \ref{appB}.

As expected, $T_{xz}=T_{zx}=T_{yz}=T_{zy}=0$, in agreement with the symmetry of the problem. As a consequence,
there are five relevant elements: the three diagonal ($T_{xx}$, $T_{yy}$, and $T_{zz}$) and two off-diagonal
elements ($T_{xy}$ and $T_{yx}$). In addition, $T_{xx}\neq T_{yy}=T_{zz}$ and $T_{xy}\neq T_{yx}$. The equality
$P_{s,yy}=P_{s,zz}$ implies $T_{yy}=T_{zz}$. This property is a consequence of the interaction model considered
since for non-Maxwell molecules computer simulations show that the $yy$ and $zz$ elements of the pressure tensor
are different. \cite{MSG96} In Figs.\ \ref{fig1}--\ref{fig6}, the relevant elements of tensors $D_{ij}^*$,
$D_{p,ij}^*$ and $D_{T,ij}^*$ are plotted as functions of the reduced shear rate $a^*$ for
$\kappa_{12}=\kappa_{22}$ and several values of the mass ratio $\mu$. Here, the tensors have been reduced with
respect to their Navier-Stokes values (except $D_{T,ij}$), namely, $D_{ij}^*=D_{ij}/D_0$,
$D_{p,ij}^*=D_{p,ij}/D_{p,0}$ and $D_{T,ij}^*=D_{T,ij}/x_1D_0$. One third of the trace of these tensors is
plotted in Figs.\ \ref{fig1}--\ref{fig3}, while the $xy$ element  is plotted in Figs. \ref{fig4}--\ref{fig6}. We
observe that in general the influence of shear flow on the mass transport is quite important. It is also
apparent that the anisotropy of the system, as measured by the traces $\frac{1}{3}D_{kk}^*$,
$\frac{1}{3}D_{p,kk}^*$, and $\frac{1}{3}D_{pTkk}^*$, grows with the shear rate. This anisotropy is more
significant when the impurity is heavier than the particles of the gas. Moreover, the shear field induces cross
effects in the diffusion of particles. This is measured by the (reduced) off-diagonal elements $D_{xy}^*$,
$D_{p,xy}^*$ and $D_{T,xy}^*$. These coefficients give the transport of mass along the $x$ axis due to gradients
parallel to the $y$ axis. While $D_{xy}^*$ and $D_{p,xy}^*$ are negative, the coefficient $D_{T,xy}$ can be
positive in the region of small shear rates. We observe that, regardless of the mass ratio, the shapes of
$D_{xy}^*$ and $D_{p,xy}^*$ are quite similar: there is a region of values of $a^*$ for which $-D_{xy}^*$ and
$-D_{p,xy}^*$ increase with increasing shear rate, while the opposite happens for larger shear rates. The
magnitude of $D_{T,xy}^*$ is smaller than that of the elements $-D_{xy}^*$ and $-D_{p,xy}^*$, especially when
the tracer particles are lighter than the particles of the gas. In this latter case, $D_{T,xy}^*$ is practically
negligible.

\section{Discussion}
\label{sec6}

Diffusion of particles in a binary mixture in non-Newtonian regimes is a subject of great interest from a
fundamental and practical points of view. If the mixture is strongly sheared, the mass flux ${\bf j}_1$ can be
significantly affected by the presence of shear flow so that the corresponding transport coefficients may differ
significantly from their equilibrium values. In addition, the resulting mass transport is anisotropic and thus
it cannot be described by scalar transport coefficients but by shear-rate dependent tensorial quantities whose
explicit determination has been the main objective of this paper.

In order to gain some insight into this complex problem, a dilute binary mixture of Maxwell molecules under USF
has been considered. This is perhaps the only interaction potential for which the Boltzmann equation can be
exactly solved in some specific non-homogenous situations, such as in the case of the USF problem. In
particular, the corresponding rheological properties of the mixture (nonlinear shear viscosity and viscometric
functions) have been obtained  for {\em arbitrary} values of the shear rate and without any restriction on the
parameters of the mixture (masses, concentrations, and force constants). \cite{MGS95} This exact solution is of
great significance in providing insight into the type of phenomena that can occur in conditions far away from
equilibrium. In this paper, the interest has been focused on situations that slightly deviate from the USF by
small spatial gradients. Under these conditions, a generalized Chapman-Enskog method \cite{LD97,L06,G06,G07}
around the shear flow distribution has been used to determine mass transport in the first order of the
deviations of the hydrodynamic field gradients from their values in the reference shear flow state $f_s^{(0)}$.
In this case, the mass flux ${\bf j}_1^{(1)}$ is given by Eq.\ (\ref{3.25}), where the corresponding set of
generalized transport coefficients $\{D_{ij}, D_{p,ij}, D_{T,ij}\}$ are the solutions of the coupled algebraic
equations (\ref{4.23})--(\ref{4.25}) in the unthermostatted case, while they are explicitly given by Eqs.\
(\ref{4.30})--(\ref{4.32}) in the presence of a Gaussian thermostat. \cite{EM90} This type of external forces
are usually employed in nonequilibrium molecular dynamics simulations to compensate exactly for the viscous
increase of temperature.

As expected, the results show that the coefficients $\{D_{ij}, D_{p,ij}, D_{T,ij}\}$ present a complex
dependence on the shear rate and on the masses, mole fractions, and force constants. This is clearly illustrated
in Figs.\ \ref{fig1}--\ref{fig6} for the tracer limit case ($x_1\to 0$). The deviations of $\{D_{ij}, D_{p,ij},
D_{T,ij}\}$ from their equilibrium values are basically due to three different reasons. First, the presence of
shear flow modifies the collision frequency of the conventional diffusion problem ($\rho \lambda_{12}/m_1m_2$)
by a shear-rate dependent term. Second, given that the binary mixture is in general constituted by particles
mechanically different, the reference shear flow states $f_1^{(0)}$ and $f_2^{(0)}$ are completely different.
This effect gives rise to terms proportional to $P_{1,ij}^*-(\rho_1/\rho)P_{ij}^*$. Third, in the
unthermostatted case, the generalized coefficients are coupled due to the inherent non-Newtonian features of the
USF state. Each one of the three effects is a different reflection of the extreme nonequilibrium conditions
present in the mixture.

It is apparent that the results presented here in the particular case of Maxwell molecules may be relevant for
interpreting computer simulation results. Sarman, Evans, and Baranyai \cite{SEB92} carried out time ago
molecular dynamics simulations in a strongly sheared Lennard-Jones binary mixture to evaluate the self- and
mutual-diffusion tensor by means of Green-Kubo formulae. \cite{E91} To the best of my knowledge, this is the
only computer experiment in which the shear-rate dependence of the diffusion tensor $D_{ij}$ has been measured.
They considered an equimolar Lennard-Jones mixture at two different densities and the parameters in the
potential were adjusted to model an argon-krypton mixture, which means that the two components are fairly
similar. As already said in Ref.\ \onlinecite{MG98}, when one considers this type of mixture ($x_1=0.5$,
$m_1/m_2=0.48$, $\kappa_{11}=\kappa_{22}=\kappa_{12}$) in the thermostatted case, the general qualitative
dependence of the (reduced) mutual diffusion tensor $D_{ij}(a^*)/D_0$ on the (reduced) shear rate agrees quite
well with computer simulations. Thus, theory and simulation predict that in general, the $xx$ element increases
to a maximum and then it decreases again, while the $yy$ element decreases with increasing shear rate. The
off-diagonal elements $xy$ and $yx$ are negative and their magnitude increases with $a^*$ for not very large
values of the shear rate. However, kinetic theory predicts that $|D_{xy}|>|D_{yx}|$, while the opposite happens
in computer simulations. On the other hand, at a quantitative level, the influence of shear flow on diffusion is
much more modest in the molecular dynamics simulations than the one found theoretically for dilute gases. This
is probably due to the fact that the shear rates (in reduced units) applied in the simulations are not large
enough to observe significant changes of the diffusion tensor relative to its equilibrium value. An alternative
to overcome the difficulties for reaching large shear rates in nonequilibrium molecular dynamics at low-density
is the direct simulation Monte Carlo method. \cite{B94} I hope that the results derived here for Maxwell
molecules for $D_{ij}$, $D_{p,ij}$ and $D_{T,ij}$ stimulate the performance of Monte Carlo simulations to assess
the reliability of the Maxwell results to describe mass transport in strongly sheared mixtures for more
realistic interaction potentials.

As said before, it must noted that, in order to observe large effects of shear flow on the tensors $\{D_{ij},
D_{p,ij}, D_{T,ij}\}$, the (reduced) shear rate must be at least of the order of 1. This means that for the
inert gas fluids considered in this paper, non-Newtonian effects on mass transport could be observable for shear
rates practically unattainable in the laboratory. In this sense, one should look at fluids that are observed to
be non-Newtonian, such as colloidal suspensions, polymeric liquids, gels,$\cdots$. \cite{B84a,B84b}

Although the results derived in this paper have been focused on the mass transport, the remaining transport
coefficients associated with the pressure tensor $P_{ij}^{(1)}$ and the heat flux vector ${\bf q}^{(1)}$ could
be determined from the integral equations (\ref{3.18})--(\ref{3.21}). Nevertheless, in practice this calculation
cannot be carried out analytically by using the Boltzmann equation since the fourth-degree moments of USF (whose
explicit expressions are not known in the Boltzmann equation, except for a single gas\cite{GS03}) are needed to
get the heat flux. In order to overcome such a difficulty one can use a model kinetic equation that preserves
the essential features of the true Boltzmann equation but admits a more practical analysis. Perhaps the most
well-known model for gas mixtures is the Gross-Krook (GK) kinetic model. \cite{GK56} In this model the Boltzmann
operator $J_{rs}[f_r,f_s]$ is replaced by the relaxation term
\begin{equation}
\label{5.1} J_{rs}[f_r,f_s]\to -\nu_{rs}(f_r-f_{rs}),
\end{equation}
where
\begin{equation}
f_{rs}=n_{r}\left( \frac{m_{r}}{2\pi k_{B}T_{rs}}\right) ^{3/2}\exp \left[ - \frac{m_{r}}{2k_{B}T_{rs}}({\bf
v}-{\bf u}_{rs})^{2}\right]  \label{5.2}
\end{equation}
and
\begin{equation}
{\bf u}_{rs}=\frac{m_{r}{\bf u}_{r}+m_{s}{\bf u}_{s}}{m_{r}+m_{s}}\;, \label{5.3}
\end{equation}
\begin{equation}
T_{rs}=T_{r}+2\frac{m_{r}m_{s}}{(m_{r}+m_{s})^{2}}\left[ (T_{s}-T_{r})+\frac{ m_{s}}{6k_{B}}({\bf u}_{r}-{\bf
u}_{s})^{2}\right] \;.  \label{5.4}
\end{equation}
The partial temperatures $T_r$ are defined by Eq.\ (\ref{2.5}). For Maxwell molecules, the effective collision
frequency $\nu_{rs}$ is given by
\begin{equation}
\nu _{rs}=An_{s}\left( \kappa _{rs}\frac{m_{r}+m_{s}}{m_{r}m_{s}}\right) ^{1/2}\;,  \label{5.5}
\end{equation}
where $A$ is a constant to be fixed by requiring that the model reproduces some transport coefficient of the
Boltzmann equation. An exact solution to the GK kinetic model for a binary mixture in USF has been found.
\cite{MG96} The comparison of the GK results with those from the Boltzmann equation \cite{MGS95} at the level of
the rheological properties shows good agreement, confirming the reliability of the GK model in computing
transport properties in far from equilibrium situations as well. Starting from the USF solution of the GK model,
the fluxes $P_{ij}^{(1)}$ and ${\bf q}^{(1)}$ are obtained in Appendix \ref{appD} in the thermostatted case.
With all the transport coefficients known, the constitutive equations for the mass, momentum and heat fluxes are
completed and the corresponding set of closed hydrodynamic equations for the mixture can be derived. This allows
one to perform a linear stability analysis of the hydrodynamic equations with respect to the USF and determine
the conditions for instabilities at long wavelengths. Previous results for a single gas \cite{LD97} have shown
that USF is unstable when the perturbations are along the velocity gradient ($y$ direction). The problem now is
to extend this analysis to the case of multicomponent systems. Work along this line will be reported in the near
future.

\acknowledgments

Partial support from the Ministerio de Ciencia y Tecnolog\'{\i}a (Spain) through Grant No. FIS2007--60977 and
from the Junta de Extremadura through Grant No. GRU07046 is acknowledged.

\appendix
\section{Chapman--Enskog-like expansion
\label{appA}}

In this Appendix, some technical details on the determination of the first-order approximation $f_1^{(1)}$ by
means of the Chapman--Enskog-like expansion are provided. Inserting the expansions (\ref{3.8}) and (\ref{3.11})
into Eq.\ (\ref{3.1}), one gets the kinetic equation for $f_1^{(1)}$:
\begin{equation}
\label{a1} \partial_t^{(0)}f_1^{(1)}-\frac{\partial}{\partial V_i}\left(a_{ij}V_j+\alpha
V_i\right)f_1^{(1)}+\alpha \delta{\bf u}\cdot \frac{\partial f_1^{(1)}}{\partial {\bf V}} +{\cal L}_1
f_1^{(1)}+{\cal M}_1 f_2^{(1)} =-\left[\partial_t^{(1)}+({\bf V}+{\bf u}_0)\cdot \nabla \right]f_1^{(1)}.
\end{equation}
The velocity dependence on the right-hand side of Eq.\ (\ref{a1}) can be obtained from the macroscopic balance
equations (\ref{3.3})--(\ref{3.5}) to first order in the gradients. They are given by
\begin{equation}
\label{a2}
\partial_t^{(1)}x_1=-({\bf u}_0+\delta {\bf u})\cdot \nabla x_1,
\end{equation}
\begin{equation}
\label{a3}
\partial_t^{(1)}\delta {\bf u}=-({\bf u}_0+\delta {\bf u})\cdot \nabla \delta {\bf u}
-{\rho}^{-1}\nabla \cdot {\sf P}^{(0)},
\end{equation}
\begin{equation}
\label{a4} \partial_t^{(1)}p=-({\bf u}_0+ \delta {\bf u})\cdot \nabla p-p\nabla \cdot \delta {\bf
u}-\frac{2}{3}\left( aP_{xy}^{(1)}+{\sf P}^{(0)}:\nabla \delta {\bf u}\right),
\end{equation}
\begin{equation}
\label{a5} \partial_t^{(1)}T=-({\bf u}_0+ \delta {\bf u})\cdot \nabla T-\frac{2}{3 n}\left( aP_{xy}^{(1)}+{\sf
P}^{(0)}:\nabla \delta {\bf u}\right),
\end{equation}
where use has been made of the result ${\bf j}_1^{(0)}={\bf q}^{(0)}={\bf 0}$. In addition,
\begin{equation}
\label{a6} P_{ij}^{(1)}=\sum_{s}\;m_s\int d{\bf c}\, c_i c_j  f_s^{(1)}({\bf c}).
\end{equation}
Use of Eqs.\ (\ref{a2})--(\ref{a5}) in Eq.\ (\ref{a1}) yields
\begin{eqnarray}
\label{a7} \partial_t^{(0)}f_1^{(1)}-\frac{\partial}{\partial V_i}\left(a_{ij}V_j+\alpha
V_i\right)f_1^{(1)}&+&\alpha \delta{\bf u}\cdot \frac{\partial f_1^{(1)}}{\partial {\bf V}} +{\cal L}_1
f_1^{(1)}+{\cal M}_1 f_2^{(1)} = {\bf A}_1\cdot \nabla x_1\nonumber\\
& & +{\bf B}_1\cdot \nabla p+{\bf C}_1\cdot \nabla T+{\sf D}_1:\nabla \delta {\bf u},
\end{eqnarray}
where
\begin{equation}
\label{a8} A_{1,i}({\bf c})=-\frac{\partial f_1^{(0)}}{\partial x_1}c_i+\frac{1}{\rho} \frac{\partial
f_1^{(0)}}{\partial \delta u_j}\frac{\partial P_{ij}^{(0)}}{\partial x_1},
\end{equation}
\begin{equation}
\label{a9} B_{1,i}({\bf c})=-\frac{\partial f_1^{(0)}}{\partial p} c_i+\frac{1}{\rho} \frac{\partial
f_1^{(0)}}{\partial \delta u_j}\frac{\partial P_{ij}^{(0)}}{\partial p},
\end{equation}
\begin{equation}
\label{a10} C_{1,i}({\bf c})=-\frac{\partial f_1^{(0)}}{\partial T} c_i+\frac{1}{\rho} \frac{\partial
f_1^{(0)}}{\partial \delta u_j}\frac{\partial P_{ij}^{(0)}}{\partial T},
\end{equation}
\begin{equation}
\label{a11} D_{1,ij}({\bf c})=p\frac{\partial f_1^{(0)}}{\partial p}\delta_{ij}-\frac{\partial f_1^{(0)}}
{\partial \delta u_i}c_j+\frac{2}{3p}\left(P_{ij}^{(0)}-a\eta_{xyij}\right) \left(p\frac{\partial}{\partial
p}+T\frac{\partial}{\partial T}\right)f_1^{(0)}.
\end{equation}
Upon writing Eq.\ (\ref{a11}) use has been made of the expression of the total pressure tensor $P_{ij}^{(1)}$ of
the mixture \cite{LD97,GS03}
\begin{equation}
\label{a12} P_{ij}^{(1)}=-\eta_{ijk\ell} \frac{\partial \delta u_k} {\partial r_{\ell}},
\end{equation}
where $\eta_{ijk\ell}$ is the viscosity tensor.

The solution to Eq.\ (\ref{a7}) has the form given by Eq.\ (\ref{3.17}), where the coefficients ${\boldsymbol
{\cal A}}_{1}$, ${\boldsymbol {\cal B}}_{1}$, ${\boldsymbol {\cal C}}_{1}$, and ${\sf {\cal D}}_{1}$ are
functions of the peculiar velocity and the hydrodynamic fields $x_1$, $p$, $T$, and $\delta {\bf u}$. The time
derivative acting on these quantities can be evaluated with the replacement
\begin{equation}
\label{a14} \partial_t^{(0)}\to -\left(\frac{2}{3 p}a
P_{xy}^{(0)}+2\alpha\right)\left(p\partial_p+T\partial_T\right).
\end{equation}
Moreover, there are contributions from $\partial_t^{(0)}$ acting on the pressure, temperature, and velocity
gradients given by
\begin{eqnarray}
\label{a15}
\partial_t^{(0)} \nabla p&=&-\nabla \left(\frac{2}{3}a
P_{xy}^{(0)}+2p\alpha\right)
\nonumber\\
&=&-\left(\frac{2a}{3}\frac{\partial P_{xy}^{(0)}}{\partial x_1}+2p\frac{\partial \alpha}{\partial
x_1}\right)\nabla x_1- \left(\frac{2a}{3} \frac{\partial P_{xy}^{(0)}}{\partial p}+ 2\alpha+2\frac{\partial
\alpha}{\partial p}\right) \nabla p \nonumber\\
& &-\left(\frac{2a}{3} \frac{\partial P_{xy}^{(0)}}{\partial T}+2p\frac{\partial \alpha}{\partial T}\right)
\nabla T,
\end{eqnarray}
\begin{eqnarray}
\label{a16}
\partial_t^{(0)} \nabla T&=&-\nabla \left(\frac{2T}{3p}a
P_{xy}^{(0)}+2T\alpha\right)
\nonumber\\
&=&-\left(\frac{2aT}{3p}\frac{\partial P_{xy}^{(0)}}{\partial x_1}+2T\frac{\partial \alpha}{\partial
x_1}\right)\nabla x_1+ \left(\frac{2aT}{3p^2}P_{xy}^{(0)}- \frac{2aT}{3p}\frac{\partial P_{xy}^{(0)}}{\partial
p}- 2T\frac{\partial
\alpha}{\partial p}\right) \nabla p \nonumber\\
& &-\left(\frac{2a}{3p}P_{xy}^{(0)}+\frac{2aT}{3p} \frac{\partial P_{xy}^{(0)}}{\partial
T}+2\alpha+2T\frac{\partial \alpha}{\partial T}\right) \nabla T,
\end{eqnarray}
\begin{equation}
\label{a17}
\partial_t^{(0)} \nabla_i \delta u_j=\nabla_i \partial_t^{(0)} \delta u_j=-a_{jk} \nabla_i \delta u_k.
\end{equation}
The corresponding integral equations (\ref{3.18})--(\ref{3.20}) can be obtained when one identifies coefficients
of independent gradients in Eq.\ (\ref{a7}) and takes into account Eqs.\ (\ref{a15})--(\ref{a17}) and the
mathematical property
\begin{eqnarray}
\label{a18}
\partial_t^{(0)} X &=&\frac{\partial X}{\partial p}\partial_t^{(0)}
p+\frac{\partial X}{\partial T}\partial_t^{(0)} T+\frac{\partial X}{\partial \delta u_i}\partial_t^{(0)}
\delta u_i \nonumber\\
&=&-\left(\frac{2}{3 p}a P_{xy}^{(0)}+2\alpha\right)\left(p\frac{\partial}{\partial p}+T\frac{\partial}{\partial
T}\right) X+a_{ij}\delta u_j \frac{\partial X}{\partial c_i},
\end{eqnarray}
where in the last step it has been taken into account that $X$ depends on $\delta {\bf u}$ through ${\bf c}={\bf
V}-\delta {\bf u}$.

\section{Generalized transport coefficients associated with the mass transport}
\label{appC}

In the unthermostatted case ($\alpha=0$), the integral equations defining the generalized transport coefficients
$D_{ij}$, $D_{p,ij}$ and $D_{T,ij}$ are given by Eqs.\ (\ref{4.1})--(\ref{4.3}). To get these coefficients, one
multiplies (\ref{4.1})--(\ref{4.3}) by $m_1 c_i$ and integrates over velocity. The result is
\begin{eqnarray}
\label{4.7}  & & \frac{2}{3 p}a
P_{xy}^{(0)}\left(p\partial_p+T\partial_T\right)\left(\frac{m_1m_2n}{\rho}D_{ij}\right)- \frac{m_1m_2n}{\rho}
\left(a_{ik}D_{kj}+\frac{\rho
\lambda_{12}}{m_1m_2}D_{ij}\right)\nonumber\\
& & =m_1\int\dd{\bf c}\; c_i A_{1,j} -\frac{2a\rho}{3p}(\partial_{x_1}P_{xy}^{(0)})
\left(D_{p,ij}+D_{T,ij}\right),
\end{eqnarray}
\begin{eqnarray}
\label{4.8} && \frac{2}{3 p}a
P_{xy}^{(0)}\left(p\partial_p+T\partial_T\right)\left(\frac{\rho}{p}D_{p,ij}\right)-
\frac{\rho}{p}\left[a_{ik}D_{p,kj}+ \left(\frac{\rho\lambda_{12}}{m_1m_2}-\frac{2a}{3 p}\partial_{p}P_{xy}^{(0)}
\right)D_{p,ij}\right]\nonumber\\
& & =m_1\int\dd{\bf c}\; c_i B_{1,j}+ \frac{2a\rho}{3p^2}D_{T,ij}(1-p\partial_p)P_{xy}^{(0)},
\end{eqnarray}
\begin{eqnarray}
\label{4.9} && \frac{2}{3 p}a
P_{xy}^{(0)}\left(p\partial_p+T\partial_T\right)\left(\frac{\rho}{T}D_{T,ij}\right)-
\frac{\rho}{T}\left[a_{ik}D_{T,kj}+ \left(\frac{\rho \lambda_{12}}{m_1m_2}-\frac{2a}{3
p}(1+T\partial_{T})P_{xy}^{(0)}\right) D_{T,ij}\right]\nonumber\\
& & =m_1\int\dd{\bf c}\; c_i C_{1,j}- \frac{2a\rho}{3p}D_{p,ij}(\partial_TP_{xy}^{(0)}),
\end{eqnarray}
where
\begin{equation}
\label{4.10} P_{s,ij}^{(0)}=m_s\int \dd{\bf c}\; c_i c_j f_s^{(0)}.
\end{equation}
Upon writing Eqs.\ (\ref{4.7})--(\ref{4.9}), use has been made of the relation (\ref{2.10.1}), which yields the
results
\begin{subequations}
\begin{equation}
\label{4.12} m_1\int \dd{\bf c}\; c_i\left({\cal L}_1{\boldsymbol {\cal A}}_{1}+{\cal M}_1{\boldsymbol {\cal
A}}_{2}\right)=-n\lambda_{12}D_{ij},
\end{equation}
\begin{equation}
\label{4.13} m_1\int \dd{\bf c}\; c_i\left({\cal L}_1{\boldsymbol {\cal B}}_{1}+{\cal M}_1{\boldsymbol {\cal
B}}_{2}\right)=-\frac{\rho^2\lambda_{12}}{m_1m_2p}D_{p,ij},
\end{equation}
\begin{equation}
\label{4.14} m_1\int \dd{\bf c}\; c_i\left({\cal L}_1{\boldsymbol {\cal C}}_{1}+{\cal M}_1{\boldsymbol {\cal
C}}_{2}\right)=-\frac{\rho^2\lambda_{12}}{m_1m_2T}D_{T,ij}.
\end{equation}
\end{subequations}
The velocity integrals appearing in Eqs.\ (\ref{4.7})--(\ref{4.9}) can be performed by using Eqs.\ (\ref
{a8})--(\ref{a10}),
\begin{equation}
\label{4.15} m_1\int \dd{\bf c}\;c_i A_{1,j}=-\left(\partial_{x_1}P_{1,ij}^{(0)}-\frac{\rho_1}{\rho}
\partial_{x_1}P_{ij}^{(0)}\right),
\end{equation}
\begin{equation}
\label{4.16} m_1\int \dd{\bf c}\; c_i
B_{1,j}=-\partial_p\left(P_{1,ij}^{(0)}-\frac{\rho_1}{\rho}P_{ij}^{(0)}\right),
\end{equation}
\begin{equation}
\label{4.17} m_1\int \dd{\bf c}\;c_i
C_{1,j}=-\partial_T\left(P_{1,ij}^{(0)}-\frac{\rho_1}{\rho}P_{ij}^{(0)}\right).
\end{equation}
The generalized transport coefficients $D_{ij}$, $D_{p,ij}$, and $D_{T,ij}$ can be written as $D_{ij}=D_0
D_{ij}^*(a^*)$, $D_{p,ij}=D_{p,0} D_{p,ij}^*(a^*)$, and $D_{T,ij}=D_{T,0} D_{T,ij}^*(a^*)$ where $D_{ij}^*$,
$D_{p,ij}^*$, and $D_{T,ij}^*$ are dimensionless functions of the shear rate. Moreover, from dimensional
analysis, $D_0\sim T$, $D_{p,0}\sim T^2/p$, and $D_{T,0}\sim T^2/p$. Therefore,
\begin{equation}
\label{4.19} \left(p\partial_p+T\partial_T\right)\left(\frac{m_1m_2n}{\rho}D_{ij}\right)=
\left(p\partial_p+T\partial_T\right)\left(\frac{m_1m_2n}{\rho}D_0D_{ij}^*\right)=\frac{m_1m_2n}{\rho}D_{ij},
\end{equation}
\begin{equation}
\label{4.20} \left(p\partial_p+T\partial_T\right)\left(\frac{\rho}{p}D_{p,ij}\right)=
\left(p\partial_p+T\partial_T\right)\left(\frac{\rho}{p}D_{p,0}D_{p,ij}^*\right)=0,
\end{equation}
\begin{equation}
\label{4.21} \left(p\partial_p+T\partial_T\right)\left(\frac{\rho}{T}D_{T,ij}\right)=
\left(p\partial_p+T\partial_T\right)\left(\frac{\rho}{T}D_{T,0}D_{T,ij}^*\right)=0,
\end{equation}
where use has been made of the identity
\begin{equation}
\label{4.22}\left(p\partial_p+T\partial_T\right)X(a^*)=(\partial_{a^*}X)\left(p\partial_p a^*+T\partial_T
a^*\right)=0,
\end{equation}
with $a^*=a/\zeta\sim T/p$. Taking into account the above results one arrives at the set of algebraic equations
(\ref{4.23})--(\ref{4.25}).

\section{Rheological properties in the tracer limit}
\label{appB}

The explicit expressions for the pressure tensors $P_{s,ij}^*\equiv P_{s,ij}/p$ in the USF are provided in this
Appendix for the special case of tracer limit ($x_1\to 0$). The nonzero elements of $P_{2,ij}^*$ are given by
\cite{GLH91}
\begin{equation}
\label{b1} P_{2,yy}^*=P_{2,zz}^*=\frac{1}{1+2\omega \alpha^*},
\end{equation}
\begin{equation}
\label{b2} P_{2,xx}^*=\frac{1+6\omega \alpha^*}{1+2\omega \alpha^*},
\end{equation}
\begin{equation}
\label{b3} P_{2,yxy}^*=-3\frac{\alpha^*}{a^*}=-\frac{\omega a^*}{(1+2\omega \alpha^*)^2},
\end{equation}
where $a^*=a/\zeta$, $\alpha^*=\alpha/\zeta$, $\zeta$ being defined by Eq.\ (\ref{4.4}). Moreover,
\begin{equation}
\label{b4} \omega=\frac{2}{\gamma_{22}(1+\mu)},\quad
\gamma_{22}=\sqrt{\frac{\kappa_{22}}{\kappa_{12}}\frac{1+\mu}{\mu}},
\end{equation}
where $\mu=m_1/m_2$ is the mass ratio. The (reduced) thermostat parameter is given by
$\alpha^*=\text{max}(\alpha_0,\alpha_0')$  where \cite{MGS95,MSG96a}
\begin{equation}
\label{b5} \alpha_0=\frac{1}{2\omega}G(\omega a^*), \quad \alpha_0'=\frac{1}{4\mu}G(2 \mu
a^*)-\frac{1}{2}\gamma_{12},
\end{equation}
where $G(z)=\frac{4}{3}\sinh^2[\frac{1}{6}\cosh^{-1}(1+9z^2)]$ and
$\gamma_{12}=\lambda_{12}/\lambda_{12}'=0.648$. Usually, $\alpha_0>\alpha_0'$ except for very large shear rates
and/or very disparate mass binary mixtures. \cite{MSG96a}

The nonzero elements of $P_{1,ij}^*$ are given by \cite{GLH91}
\begin{eqnarray}
\label{b6} P_{1,yy}^*=P_{1,zz}^*&=&\frac{x_1}{\Delta(1+2\omega
\alpha^*)}\left\{(\gamma_{12}-\frac{1}{2})(2\alpha^*+\epsilon)^2+(2\gamma_{12}-1)\alpha^*\beta(1+\omega
\epsilon+4\omega \alpha^*)\right.
\nonumber\\
& &\left.+\frac{1}{2}(1+2\omega \alpha^*)(2\alpha^*+\epsilon)^2\right\},
\end{eqnarray}
\begin{eqnarray}
\label{b7} P_{1,xx}^*&=&\frac{x_1}{\Delta(1+2\omega
\alpha^*)}\left\{(\gamma_{12}-\frac{1}{2})(2\alpha^*+\epsilon)^2+3(2\gamma_{12}-1)\alpha^*(1+\omega
\epsilon+4\omega \alpha^*)\left(2\alpha^*+\epsilon-\frac{2}{3}\beta\right)\right.
\nonumber\\
& &\left.+\frac{1}{2}(1+2\omega
\alpha^*)\left[(2\alpha^*+\epsilon)^2+2a^{*2})+(2\gamma_{12}-1)a^{*2}\right]\right\},
\end{eqnarray}
\begin{eqnarray}
\label{b8} P_{1,xy}^*&=&-\frac{x_1\omega a^*}{\Delta(1+2\omega
\alpha^*)^2}\left\{(\gamma_{12}-\frac{1}{2})(2\alpha^*+\gamma_{12})(2\alpha^*+\epsilon)+(\gamma_{12}-\frac{1}{2})
\frac{1+2\omega
\alpha^*}{\omega}\right.\nonumber\\
& & \times \left.(2\alpha^*+\epsilon+2\omega \alpha^*\beta)+\frac{1}{2\omega}(1+2\omega
\alpha^*)^2(2\alpha^*+\epsilon)\right\},
\end{eqnarray}
where $\epsilon=\gamma_{12}+\beta$, $\beta=1/2\mu$, and
\begin{equation}
\label{b9} \Delta=(2\alpha^*+\epsilon)^2(2\alpha^*+\gamma_{12})-\frac{2}{3}\beta a^{*2}.
\end{equation}

\section{Momentum and heat transport around USF from the GK model}
\label{appD}

This Appendix addresses the evaluation of the fluxes $P_{ij}^{(1)}$ and ${\bf q}^{(1)}$ in the thermosttated
case from the GK kinetic model (\ref{5.1}). The first order corrections to the fluxes are
\begin{equation}
\label{d1} P_{ij}^{(1)}=-\sum_s\; \eta_{s,ijk\ell}\frac{\partial \delta u_k}{\partial r_{\ell}},
\end{equation}
\begin{equation}
\label{d2} q_{i}^{(1)}=-\sum_s\; D''_{s,ij}\frac{\partial x_1}{\partial r_j}-\sum_s\; L_{s,ij}\frac{\partial
p}{\partial r_j}-\sum_s\; \lambda_{s,ij}\frac{\partial T}{\partial r_j},
\end{equation}
where the partial contributions to the transport coefficients are defined as
\begin{equation}
\label{d3} \eta_{s,ijk\ell}=-m_s\int\dd{\bf c}\; c_ic_j {\cal D}_{s,k\ell}({\bf c}),
\end{equation}
\begin{equation}
\label{d4} D''_{s,ij}=-\frac{m_s}{2}\int\dd{\bf c}\; c^2c_i {\cal A}_{s,j}({\bf c}),
\end{equation}
\begin{equation}
\label{d5} L_{s,ij}=-\frac{m_s}{2}\int\dd{\bf c}\; c^2c_i {\cal B}_{s,j}({\bf c}),
\end{equation}
\begin{equation}
\label{d6} \lambda_{s,ij}=-\frac{m_s}{2}\int\dd{\bf c}\; c^2c_i {\cal C}_{s,j}({\bf c}).
\end{equation}
From the above partial contributions one can get the generalized shear viscosity
$\eta_{ijk\ell}=\eta_{1,ijk\ell}+\eta_{2,ijk\ell}$, the generalized Duffour coefficient
$D''_{ij}=D''_{1,ij}+D''_{2,ij}$, the generalized pressure energy coefficient $L_{ij}=L_{1,ij}+L_{2,ij}$, and
the generalized thermal conductivity $\lambda_{ij}=\lambda_{1,ij}+\lambda_{2,ij}$. The quantities $\{{\cal
A}_{s,i}, {\cal B}_{s,i}, {\cal C}_{s,i}, {\cal D}_{s,ij}\}$ still verify the integral equations
(\ref{3.18})--(\ref{3.21}) (with $\alpha=-aP_{xy}^{(0)}/3p$) with the only replacement
\begin{equation}
\label{d7} {\cal L}_1f_1^{(1)}+{\cal M}_1f_2^{(1)}\to \nu_1 f_1^{(1)}-\nu_{11}f_{11}^{(1)}-\nu_{12}f_{12}^{(1)},
\end{equation}
where $\nu_1=\nu_{11}+\nu_{12}$,
\begin{equation}
\label{d8} f_{11}^{(1)}=\frac{f_{11}^{(0)}}{n_1k_BT_1}{\bf c}\cdot {\bf j}_1^{(1)}, \quad
 f_{12}^{(1)}=\frac{f_{12}^{(0)}}{n_1n_2k_BT_{12}}\mu_{12}(n_2-n_1){\bf c}\cdot {\bf j}_1^{(1)}
,
 \end{equation}
 and
\begin{equation}
 f_{rs}^{(0)}=n_{r}\left( \frac{m_{r}}{2\pi k_{B}T_{rs}}\right) ^{3/2}\exp \left( -
\frac{m_{r}c^2}{2k_{B}T_{rs}}\right). \label{d9}
\end{equation}
In Eq.\ (\ref{d8}), $\mu_{rs}=m_r/(m_r+m_s)$.

In order to get the coefficients $\eta_{s,ijk\ell}$, $D''_{s,ij}$, $L_{s,ij}$ and $\lambda_{s,ij}$, it is
convenient to introduce the velocity moments
\begin{equation}
\label{d10} X_{k,\ell,m}^{(i)}=\int \dd{\bf c}\; c_x^kc_y^\ell c_z^m {\cal A}_{1,i},
\end{equation}
\begin{equation}
\label{d11} Y_{k,\ell,m}^{(i)}=\int \dd{\bf c}\; c_x^kc_y^\ell c_z^m {\cal B}_{1,i},
\end{equation}
\begin{equation}
\label{d12} Z_{k,\ell,m}^{(i)}=\int \dd{\bf c}\; c_x^kc_y^\ell c_z^m {\cal C}_{1,i},
\end{equation}
\begin{equation}
\label{d13} W_{k,\ell,m}^{(ij)}=\int \dd{\bf c}\; c_x^kc_y^\ell c_z^m {\cal D}_{1,ij},
\end{equation}
and similar definitions for the species $2$. The knowledge of the above moments allows one to get the
expressions of the coefficients $\eta_{1,ijk\ell}$, $D''_{1,ij}$, $L_{1,ij}$ and $\lambda_{1,ij}$. The method to
evaluate the moments $X_{k,\ell,m}^{(i)}$, $Y_{k,\ell,m}^{(i)}$, $Z_{k,\ell,m}^{(i)}$, and $W_{k,\ell,m}^{(ij)}$
is quite similar. Here, as an example, the mathematical steps to determine the moments $X_{k,\ell,m}^{(i)}$
associated with the transport coefficients $D''_{1,ij}$ will be analyzed in detail. First, in the thermosttated
case, Eq.\ (\ref{3.18}) with the change (\ref{d7}) becomes
\begin{equation}
\label{d14} -\left(a c_y\frac{\partial}{\partial c_x}-\nu_1+\alpha\frac{\partial}{\partial {\bf c}}\cdot {\bf
c}\right){\cal A}_{1,i}
+\frac{m_1m_2n}{\rho}\left(\frac{\nu_{11}}{n_1k_BT_1}f_{11}^{(0)}+\frac{\nu_{12}}{n_1n_2k_BT_{12}}\mu_{12}
(n_2-n_1)f_{12}^{(0)}\right)c_jD_{ji}=A_{1,i},
\end{equation}
where ${\bf A}_{1}$ is given by Eq.\ (\ref{a8}). Upon writing (\ref{d14}) use has been made of the constitutive
form (\ref{3.25}) for the mass flux. Now, we multiply Eq.\ (\ref{d14}) by $c_x^kc_y^\ell c_z^m$ and integrate
over velocity. After some algebra, we get
\begin{equation}
\label{d15} ak X_{k-1,\ell+1,m}^{(i)}+\left[\nu_1 +(k+\ell+m)\alpha\right]
X_{k,\ell,m}^{(i)}=R_{k,\ell,m}^{(i)},
\end{equation}
where
\begin{eqnarray}
\label{d16}
R_{k,\ell,m}^{(i)}&=&\overline{A}_{1,i}-\frac{m_1m_2n}{\rho}\left(\frac{2k_BT}{m_1}\right)^{(k+\ell+m+1)/2}
\left[\frac{\nu_{11}}{k_BT} \chi_1^{(k+\ell+m-1)/2}+\frac{\nu_{12}}{n_2k_BT}\mu_{12}
(n_2-n_1)\chi_{12}^{(k+\ell+m-1)/2}\right]\nonumber\\
& & \times \Lambda_{k+\delta_{jx},\ell+\delta_{jy},m+\delta_{jz}}D_{ji},
\end{eqnarray}
\begin{eqnarray}
\label{d17} \overline{A}_{1,i}&\equiv& \int\, \dd{\bf c}\, c_x^k c_y^\ell c_z^m A_{1,i}
=-\frac{\partial}{\partial x_1}M_{k+\delta_{ix},\ell+\delta_{iy},m+\delta_{iz}} \nonumber\\
& &  +\frac{1}{\rho}\frac{\partial P_{ij}^{(0)}}{\partial x_1}\left(\delta_{jx}kM_{k-1,\ell,m}+ \delta_{jy}\ell
M_{k,\ell-1,m}+\delta_{jz}m M_{k,\ell,m-1}\right).
\end{eqnarray}
In Eqs.\ (\ref{d16}) and (\ref{d17}), we have introduced the temperature ratios $\chi_1=T_1/T$ and
$\chi_{12}=T_{12}/T$ and the (unperturbed) moments of the USF
\begin{equation}
\label{d18} M_{k,\ell,m}=\int\, \dd{\bf c}\, c_x^k c_y^\ell c_z^m f_1^{(0)}({\bf c}).
\end{equation}
The explicit shear-rate dependence of $\chi_1$, $\chi_{12}$ and $M_{k,\ell,m}$ can be found in Ref.\
\onlinecite{MG96}. Moreover,
\begin{equation}  \label{d19}
\Lambda_{k,\ell,m}=\pi^{-3/2}\Gamma (\frac{k+1}{2})\Gamma (\frac{\ell+1}{2}) \Gamma (\frac{m+1}{2})
\end{equation}
if $(k,\ell,m)$ are even, being zero otherwise. The solution to Eq.\ (\ref{d15}) can be written as
\begin{equation}
X_{k,\ell ,m}^{(i)}=\sum_{q=0}^{k}\frac{k!}{(k-q)!}(-a)^{q}[\nu _{1}+(k+\ell +m)\alpha ]^{-(1+q)}R_{k-q,\ell
+q,m}^{(i)}\;.  \label{d20}
\end{equation}
Note that Eq.\ (\ref{d20}) is still formal since one needs to know the coefficients $D_{ij}$. They can be
consistently determined from their definitions (\ref{3.26}). Once these coefficients are known, Eq.\ (\ref{d20})
allows one to get the coefficients $D''_{1,ij}$.

The same method can be applied to evaluate the remaining moments. The moments $Y_{k,\ell ,m}^{(i)}$ and
$Z_{k,\ell ,m}^{(i)}$ are given by
\begin{equation}
Y_{k,\ell ,m}^{(i)}=\sum_{q=0}^{k}\frac{k!}{(k-q)!}(-a)^{q}[\nu _{1}+(k+\ell +m)\alpha ]^{-(1+q)}S_{k-q,\ell
+q,m}^{(i)}\;,  \label{d21}
\end{equation}
\begin{equation}
Z_{k,\ell ,m}^{(i)}=\sum_{q=0}^{k}\frac{k!}{(k-q)!}(-a)^{q}[\nu _{1}+(k+\ell +m)\alpha ]^{-(1+q)}
T_{k-q,\ell+q,m}^{(i)}\;,  \label{d21.1}
\end{equation}
where
\begin{eqnarray}
\label{d22}
S_{k,\ell,m}^{(i)}&=&\overline{B}_{1,i}-\frac{\rho}{p}\left(\frac{2k_BT}{m_1}\right)^{(k+\ell+m+1)/2}
\left[\frac{\nu_{11}}{k_BT} \chi_1^{(k+\ell+m-1)/2}+\frac{\nu_{12}}{n_2k_BT}\mu_{12}
(n_2-n_1)\chi_{12}^{(k+\ell+m-1)/2}\right]\nonumber\\
& & \times \Lambda_{k+\delta_{jx},\ell+\delta_{jy},m+\delta_{jz}}D_{p,ji},
\end{eqnarray}
\begin{eqnarray}
\label{d23}
T_{k,\ell,m}^{(i)}&=&\overline{C}_{1,i}-\frac{\rho}{T}\left(\frac{2k_BT}{m_1}\right)^{(k+\ell+m+1)/2}
\left[\frac{\nu_{11}}{k_BT} \chi_1^{(k+\ell+m-1)/2}+\frac{\nu_{12}}{n_2k_BT}\mu_{12}
(n_2-n_1)\chi_{12}^{(k+\ell+m-1)/2}\right]\nonumber\\
& & \times \Lambda_{k+\delta_{jx},\ell+\delta_{jy},m+\delta_{jz}}D_{T,ji}.
\end{eqnarray}
The expressions of $\overline{B}_{1,i}$ and $\overline{C}_{1,i}$ are formally identical to that of
$\overline{A}_{1,i}$, except that the operator $\partial_{x_1}$ appearing in (\ref{d17}) must be replaced by the
operators $\partial_p$ and $\partial_T$ in the cases of $\overline{B}_{1,i}$ and $\overline{C}_{1,i}$,
respectively. Finally, the expression of $W_{k,\ell ,m}^{(ij)}$ is
\begin{equation}
\label{d22bis} W_{k,\ell ,m}^{(ij)}=\sum_{q=0}^{k}\frac{k!}{(k-q)!}(-a)^{q}[\nu _{1}+(k+\ell +m)\alpha
]^{-(1+q)} \left[ U_{k-q,\ell+q,m}^{(ij)}+a\delta_{iy}W_{k,\ell ,m}^{(xj)}\right],
\end{equation}
where
\begin{eqnarray}
\label{d23bis} U_{k,\ell,m}^{(ij)}&=& -\delta_{ij}\left(1-p\frac{\partial}{\partial
p}\right)M_{k,\ell,m}+\frac{2}{3p} \left(P_{ij}^{(0)}-a\eta_{xyij}\right)
\left(p\frac{\partial}{\partial p}+T\frac{\partial}{\partial T}\right)M_{k,\ell,m}\nonumber\\
& & -M_{k,\ell,m}\left(\delta_{ix}\delta_{j x}k+\delta_{iy}
\delta_{j y}\ell+\delta_{iz}\delta_{j z}m\right)\nonumber\\
& & -k\delta_{ix}\left(\delta_{j y}M_{k-1,\ell+1,m}+
\delta_{j z}M_{k-1,\ell,m+1}\right)\nonumber\\
& & -\ell\delta_{iy}\left(\delta_{j x}M_{k+1,\ell-1,m}+
\delta_{j z}M_{k,\ell-1,m+1}\right)\nonumber\\
& & -m\delta_{iz}\left(\delta_{j x}M_{k+1,\ell,m-1}+ \delta_{j y}M_{k,\ell+1,m-1}\right).
\end{eqnarray}

The transport coefficients $D''_{1,ij}$, $L_{1,ij}$, $\lambda_{1,ij}$, and $\eta_{1,ijk\ell}$ can be obtained
from Eqs.\ (\ref{d20}), (\ref{d21}), (\ref{d21.1}) and (\ref{d22bis}), respectively, in terms of the shear rate
and the parameters of the mixture. Their respective counterparts for species $2$ can be easily determined from
them by making the changes: $m_1\to m_2$, $n_1\to n_2$, and $\kappa_{11}\to \kappa_{22}$. The expression of the
Duffour tensor $D_{ij}''$ coincides with the one obtained before \cite{MGLH00} in a stationary state with
$\nabla p=\nabla T=0$. Finally, note that the USF moments $M_{k,\ell,m}$ can be written as \cite{MG96}
\begin{equation}
\label{d24} M_{k,\ell,m}=n_1\left(\frac{2k_BT}{m_1}\right)^{(k+\ell+m)/2}M_{k,\ell,m}^*,
\end{equation}
where the dimensionless moments $M_{k,\ell,m}^*$ depend on $p$ and $T$ through their dependence on $a^*$.
Consequently,
\begin{equation}
\label{d25} p\partial_p M_{k,\ell,m}=n_1\left(\frac{2k_BT}{m_1}\right)^{(k+\ell+m)/2}\left(1-a^*\partial_{a^*}
\right)M_{k,\ell,m}^*,
\end{equation}
\begin{equation}
\label{d26} T\partial_T
M_{k,\ell,m}=n_1\left(\frac{2k_BT}{m_1}\right)^{(k+\ell+m)/2}\left(\frac{k+\ell+m-2}{2}+a^*\partial_{a^*}
\right)M_{k,\ell,m}^*.
\end{equation}

\end{document}